\documentclass[aps,prb,reprint,floatfix]{revtex4-1}
\usepackage{amsmath,amsfonts,amssymb} %
\usepackage{mathtools} %
\usepackage{wasysym} %
\usepackage{bm} %
\usepackage[bbgreekl]{mathbbol} %
\usepackage{graphicx} %
\usepackage[dvipsnames,table]{xcolor} %
\usepackage{tabularx} %
\usepackage[acronym]{glossaries} %
\usepackage{braket} %
\usepackage{multirow} % Required for multicells

\usepackage{fancyhdr} %
\usepackage{microtype} %
\usepackage{natmove}

\usepackage[byname]{smartref} %
\usepackage[breaklinks, %
colorlinks, linkcolor=Blue, citecolor=Blue, urlcolor=Blue]{hyperref} %
\usepackage[version=3]{mhchem}
\AtBeginDocument{\usepackage{booktabs}} % Fix incompatibility with

\newcommand{\RR}{\mathbf{R}} %
\newcommand{\BC}{\color{black} } %
\newcommand{\eq}[1]{Eq.~(\ref{#1})} %
\newcommand{\fig}[1]{Fig.~\ref{#1}} %

\def\be{\begin{equation}} %
\def\ee{\end{equation}} %
\def\bea{\begin{eqnarray}} %
\def\eea{\end{eqnarray}} %

\newacronym{QPE}{QPE}{quantum phase estimation} %
\newacronym{VQE}{VQE}{Variational Quantum Eigensolver} %
\newacronym{UCC}{UCC}{unitary coupled cluster} %
\newacronym{QCC}{QCC}{qubit coupled cluster} %
\newacronym{FCI}{FCI}{full configurational interaction} %
\newacronym{CASCI}{CASCI}{complete active space configurational
  interaction} %
\newacronym{JW}{JW}{Jordan--Wigner} %
\newacronym{BK}{BK}{Bravyi--Kitaev} %
\newacronym[longplural={degrees of freedom}, %
firstplural={degrees of freedom (DOF)}, plural={DOF}]{DOF}{DOF}{degree
  of freedom} %
\newacronym[longplural={equations of motion}, %
firstplural={equations of motion (EOM)}, %
plural={EOM}]{EOM}{EOM}{equation of motion} %
\newacronym{PES}{PES}{potential energy surface} %
\newacronym{CI}{CI}{configuration interaction} %
\newacronym{QMF}{QMF}{qubit mean-field} %
\newacronym{SQP}{SQP}{sequential quadratic programming} %
\newacronym{RHF}{RHF}{restricted Hartree--Fock}

\begin{document}

\title{Measuring all compatible operators in one series of single-qubit measurements using unitary transformations}

\author{Tzu-Ching Yen${}^a$} 
\affiliation{${}^a$Chemical Physics Theory Group, Department of Chemistry,
  University of Toronto, Toronto, Ontario, M5S 3H6, Canada; 
  ${}^b$ Department of Physical and Environmental Sciences,
  University of Toronto Scarborough, Toronto, Ontario, M1C 1A4,
  Canada; ${}^c$Department of Quantum Field Theory, Taras Shevchenko National University of Kyiv, Kyiv, 03022, Ukraine}

\author{Vladyslav Verteletskyi${}^{a,b,c}$} 
\affiliation{${}^a$Chemical Physics Theory Group, Department of Chemistry,
  University of Toronto, Toronto, Ontario, M5S 3H6, Canada; 
  ${}^b$ Department of Physical and Environmental Sciences,
  University of Toronto Scarborough, Toronto, Ontario, M1C 1A4,
  Canada; ${}^c$Department of Quantum Field Theory, Taras Shevchenko National University of Kyiv, Kyiv, 03022, Ukraine}
  
\author{Artur F. Izmaylov${}^{a,b}$} 
\affiliation{${}^a$Chemical Physics Theory Group, Department of Chemistry,
  University of Toronto, Toronto, Ontario, M5S 3H6, Canada; 
  ${}^b$ Department of Physical and Environmental Sciences,
  University of Toronto Scarborough, Toronto, Ontario, M1C 1A4,
  Canada; ${}^c$Department of Quantum Field Theory, Taras Shevchenko National University of Kyiv, Kyiv, 03022, Ukraine}
  \email{artur.izmaylov@utoronto.ca}

\begin{abstract}
The Variational Quantum Eigensolver approach to the electronic structure problem on a quantum 
computer involves measurement of the Hamiltonian expectation value. 
Formally, quantum mechanics allows one to measure all mutually commuting or compatible operators simultaneously.
Unfortunately, the current hardware permits measuring only a much more limited subset of operators that 
share a common tensor product eigen-basis. 
We introduce unitary transformations that transform any fully commuting group of operators 
to a group that can be measured on current hardware. 
These unitary operations can be encoded as a sequence of Clifford gates 
and let us not only measure much larger groups of terms but also to obtain these groups efficiently 
on a classical computer. 
The problem of finding the minimum number of fully commuting groups of terms covering the whole Hamiltonian 
is found to be equivalent to the minimum clique cover problem for a graph representing Hamiltonian terms as vertices 
and commutativity between them as edges. 
Tested on a set of molecular electronic Hamiltonians with up to 50 thousand terms, 
the introduced technique allows for the reduction of the number of separately measurable operator groups  
down to few hundreds, thus achieving up to 2 orders of magnitude reduction. 
Based on the test set results, the obtained gain scales at least linearly with the number of qubits.  
\end{abstract}

\glsresetall

\maketitle

%%%%%%%%%%%%%%%%%%%%%%%%%%%%%%%%%%%%%%%%%%%%%%%%%%%%%%%%%%%%%%%%%%%%%%
\section{Introduction}

%VQE $->$ used in several areas (optimization in general, prime factorization), 
Using quantum superposition and entanglement, quantum computers promise a new powerful 
route to solve problems that are exponentially hard for their classical counterparts.
Even though quantum hardware advancements generated a surge of interest in developing new 
algorithms to solve these hard problems, we are still in the era of noisy intermediate scale quantum (NISQ)
computing.\cite{Preskill:2018jf} One of the hallmarks of NISQ algorithms is hybrid quantum-classical 
optimization of parametrized quantum circuits. The \gls{VQE} approach is one of the most popular 
realizations of this idea for solving optimization problems by mapping their solution to lowest 
eigen-states of a particular Hamiltonian with 
a bounded spectrum.\cite{Peruzzo:2014/ncomm/4213,McArdle:2018we,Cao:2018vo}  In this case, the optimization is simplified by the existence of the variational theorem 
that guarantees that any trial wavefunction will approach the solution from above. In VQE, the quantum computer 
prepares a trial wavefunction $\ket{\Psi_q}$ and 
estimates an expectation value for the target Hamiltonian $\bar{H} = \bra{\Psi_q} \hat H_q \ket{\Psi_q}$. The 
classical computer suggests a next trial wavefunction using results of expectation values based on 
previous wavefunctions. 

One of the exponentially hard and thus attractive problems that is highly valuable for chemistry
is the electronic structure problem.
\cite{AspuruGuzik:2005/sci/1704,Babbush:2014/sr/6603,Kandala:2017/nature/242,McArdle:2018we,Cao:2018vo,Genin:2019te} Its solution
provides a route to predicting chemical properties and designing many new valuable compounds such as 
materials and drugs. It is formulated using the Born-Oppenheimer separation of 
nuclear variables as parameters for the electronic part of the time-independent molecular Schrodinger 
equation
\bea
\hat H_e (\RR) \ket{\Psi(\RR)} = E_e(\RR) \ket{\Psi(\RR)},
\eea
where $\hat H_e (\RR)$ is the electronic Hamiltonian, $\RR$ is the nuclear configuration of interest, 
and $E_e(\RR)$ is the electronic energy. To treat this problem within the VQE framework
it can be mapped to a qubit eigenvalue problem 
\bea
\hat H_q (\RR)\ket{\Psi_q(\RR)} = E_e(\RR) \ket{\Psi_q(\RR)},
\eea
where $\hat H_q(\RR)$ is the qubit Hamiltonian obtained from a second quantized form of 
$\hat H_e (\RR)$\cite{Helgaker:2000} using one of the fermion-qubit mappings,\cite{Bravyi:2002/aph/210, Seeley:2012/jcp/224109,Tranter:2015/ijqc/1431, Setia:2017/ArXiv/1712.00446,Havlicek:2017/pra/032332} and $\ket{\Psi_q(\RR)}$ is the corresponding qubit wave-function. For a molecule, the qubit Hamiltonian is a linear combination 
\begin{equation}
  \label{eq:Hq}
  \hat H_q(\RR) = \sum_I C_I(\RR)\,\hat P_I
\end{equation}
of Pauli tensor products $\hat P_I$ defined as  
\begin{equation}
  \label{eq:Pi}
  \hat P_I = \prod_{i=1}^{N} \hat \sigma_{i}^{(I)},
\end{equation}
where $\hat \sigma_i^{(I)}$ is one of the $\hat x,\hat y,\hat z$ Pauli operators or the identity operator 
$\hat e$ for the $i^{\rm th}$ qubit. 
The number of qubits, $N$, is equal to the number of molecular spin-orbitals used in the second quantized 
form of the electronic Hamiltonian. Below, for simplicity, we will skip the nuclear configuration $\RR$ but always 
assume its existence as a parameter. 

%\bea
%\min_{\hat U} \bar{E}(\hat U)
%\eea

Besides problems associated with devising low-depth circuits for accurate preparation of $\ket{\Psi_q}$, the electronic 
structure problem poses another difficulty for VQE, namely estimation of the expectation value for the qubit Hamiltonian
$\hat H_q$. Note that in contrast to quantum simulators,\cite{cirac:2012,cirac:2018} in 
universal gate quantum computing, $\hat H_q$ originated 
from $\hat H_e$ is not physically implemented and does not correspond to the Hamiltonian of physical qubits. 
This makes its measurement a difficult task similar to the quantum tomography,\cite{PhysRevA.40.2847,Cramer:2010bs} 
with the only simplification that one needs to measure an $O(N^4)$ subset of the total $4^N$ set of Pauli products.
 
% EStr is one of them   - very important for applications \\
% It is the one that has relatively difficult H's to measure (intro $H_q$), not quite the tomography $4^N$ 
% but close to it $N^4$  
%
% Measurement problem - there is no Hq in the universal quantum computer which is different from the quantum simulator \\
Previously, the measurement problem has been addressed through grouping of
Pauli products whose expectation value can be 
measured simultaneously.\cite{Kandala:2017/nature/242,VVpaper1} 
Owing to additivity of the total Hamiltonian expectation value such grouping allows the
reduction of the number of separately measured operators. Considering that current hardware can only measure 
single-qubit operators and during the measurement the total wavefunction collapses to a tensor product state of 
one-qubit eigenstates of measured operators, only Pauli products that have a common 
tensor product eigen-basis (TPE) can be grouped together for simultaneous measurement.\cite{VVpaper1} 
It was found to be possible to reformulate 
optimal grouping of terms based on their shared TPE as a graph minimum clique cover (MCC) problem. 
This reformulation gave a systematic approach to reduction of the total number of terms approximately three times
from the total number of Hamiltonian terms.\cite{VVpaper1}   

TPE based grouping somewhat reduces the prefactor of the $O(N^4)$ dependence for the number of measured 
groups but does not change the scaling. Recently another approach has been put forward: 
mean-field partitioning.\cite{Izmaylov:2019gb} % and 2) partitioning based on anti-commutative 
Even though it gave some advantage compared to the TPE based partitioning, it requires  
introducing feed-forward measurement\cite{nArriagada:2018ju,Prevedel:2007ca,Moqanaki:2015iw,Reimer:2018cv} 
that has not yet become available in mainstream quantum hardware.
Also, the assessment of mean-field partitioning is hindered by the absence of an optimal partitioning algorithm.
 
%Current Advances:  MF \& AC - partitioning of Hamiltonian on MF parts or unitary parts - 
%also provides advantages on top of QWC, however, MF has no efficient partitioning and hardware to run,
%AC has a larger cost for unitaries. If one can reduce the group numbers even further without significant 
%cost of unitaries.  

Here we propose a different approach: starting with groups of fully commuting terms we convert them 
into TPE sharing groups by introducing unitary transformations. 
The necessary unitary transformations are obtained analytically using an extension of 
symplectic geometry techniques developed by Bravyi {\it et al.}\cite{Bravyi:2017wb} for tapering off qubits.
In general, sizes of fully commuting groups in qubit Hamiltonians 
are much larger than those of TPE sharing groups, and the 
proposed technique allows for the scaling reduction of the number of simultaneously measurable groups, 
which significantly increases a range of molecular systems amenable to VQE studies 
on NISQ hardware.
%Our current approach $H_q = \sum_k U_k \hat A_k U_k^\dagger $, 
%extension to even a single measurement but $U_k$ becomes too complicated

%The rest of the paper is organized as follows ...

\section{Theory}
\label{sec:theory}

\subsection{Tensor product eigen-basis sharing groups} 

More insightful and practical criterion for grouping terms sharing TPE
can be formulated using qubit-wise commutativity: two Pauli products are qubit-wise commutative if 
each single-qubit Pauli operator in one product commutes with its counterpart in the other product.
% Qubit-wise commutativity is more strict condition 
{\BC Qubit-wise commutativity is a stricter condition}
than regular commutativity and thus can be 
considered as sufficient but not necessary 
for the latter. For example, $\hat x_1 \hat x_2$ and $\hat x_1\hat e_2$ are both commutative and 
qubit-wise commutative, but 
$\hat x_1 \hat x_2$ and $\hat y_1 \hat y_2$ are only commutative. 

In the conventional VQE scheme, $\hat H_q$ is separated into sums of 
qubit-wise commuting (QWC) terms, 
\bea\label{eq:split}
\hat H_q &=&  \sum_{n=1}^{M_{\rm QWC}} \hat A_n, \\
 \hat A_n &=& \sum_{I} C_{I}^{(n)}\,\hat P_{I}^{(n)}, % ~\forall I \& J.
\eea
where all $\hat P_{I}^{(n)}$ within one $\hat A_n$ group are qubit-wise commuting.   
Partitioning of $\hat H_q$ in Eq.~(\ref{eq:split}) allows one to measure all Pauli products within each $\hat A_n$ term 
in a single set of $N$ one-qubit measurements. For every qubit, it is known from the form of
$\hat A_n$, which Pauli operator needs to be measured. The advantage of this scheme is that 
it requires only single-qubit measurements, which are technically easier than multi-qubit measurements.
Also, qubit-wise commutativity between terms provides a binary symmetric relation that is convenient for reformulation 
of optimal grouping as the MCC problem for a graph obtained by connecting 
$\hat H_q$ Pauli products (vertices) that satisfy the qubit-wise commutativity relation. 
MCC is a partitioning of the Hamiltonian graph to the minimum number of fully connected 
(complete) subgraphs or cliques. The cliques represent terms that can be measured simultaneously. 

The disadvantage of this scheme is that the Hamiltonian may require measuring too many $\hat A_n$ 
terms separately (see Ref.~\citenum{VVpaper1} for examples, typically optimal partitioning of 
$\hat H_q$ to QWC parts gives only reduction by a factor of 3 with respect to the total number of Pauli 
products in $\hat H_q$).  

On the other hand, quantum mechanics allows us to determine eigenvalues of all mutually commuting 
operators at the same time. Therefore, potentially one can partition the Hamiltonian into 
groups of fully commuting terms 
\bea\label{eq:FC}
\hat H_q &=&  \sum_{n=1}^{M_{\rm C}} \hat H_n,\\%~ [\hat H_n,\hat H_k] \ne 0,~{\rm if}~n\ne k \\
 \hat H_n &=& \sum_{I} C_{I}^{(n)}\,\hat P_{I}^{(n)}, ~ [\hat P_{I}^{(n)},\hat P_{J}^{(n)}] = 0, % ~\forall I \& J,
\eea
and measure their expectation values.
Clearly, because all QWC terms are also fully commuting the number of $\hat H_n$
groups, $M_{\rm C}$, will not be larger than that for $\hat A_n$ groups, $M_{\rm QWC}$. 
Moreover, Appendix A shows that 
the ratio between the numbers of Pauli products that commute and qubit-wise commute 
with an average Pauli product grows exponentially with the number of qubits. 
Two questions arise: 1) Is it possible 
to use the partitioning to fully commuting groups of terms in VQE without hardware modification?, and  
2) How to find the optimal partitioning of the Hamiltonian to fully commuting groups of terms?  

\subsection{Unitary Transformations}

\begin{figure}[h!]
  \centering %
  \includegraphics[width=1\columnwidth]{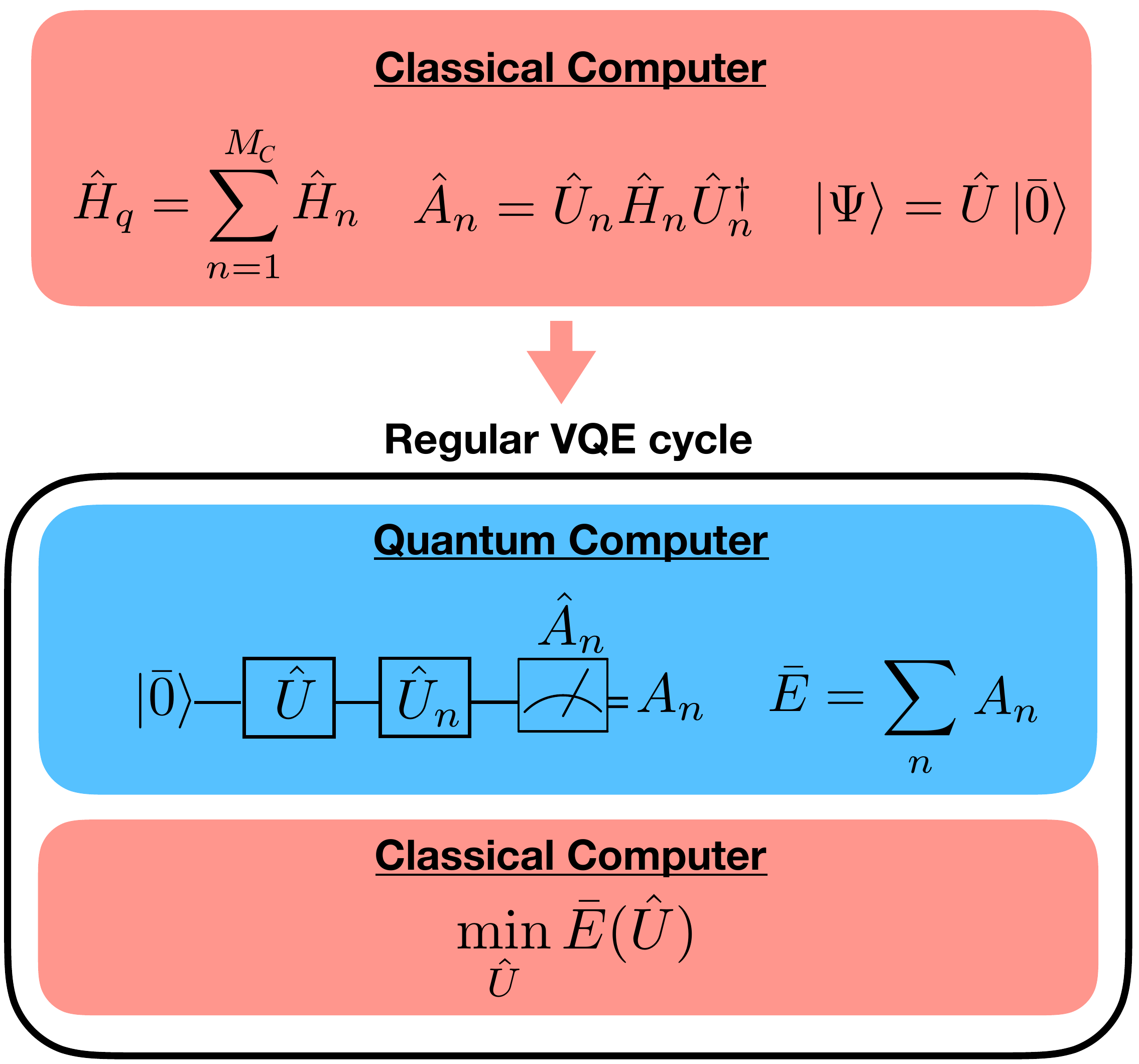}
  \caption{
      % New computational scheme.
       New computational scheme. In addition to suggesting trial wavefunctions, the classical computer identifies fully commuting groups $\hat H_n$ to be transformed into QWC groups $\hat A_n$.
  }
  \label{fig:scheme}
\end{figure}

To use more efficient partitioning in groups of fully commuting terms and keep the same single-qubit measurement 
protocol we introduce additional unitary transformations $\{\hat U_n\}$ that transform each fully commuting group 
$\hat H_n$ into a QWC group, $\hat A_n = \hat U_n \hat H_n \hat U_n^\dagger$. Note that 
$\hat A_n$ are not necessarily the same QWC operators that appear in the QWC partitioning of $\hat H_q$.
Let us consider 
partitioning of the $\hat H_q$ into fully commuting terms for the energy expectation value on a trial wavefunction 
$\ket{\Psi} = \hat U \ket{\bar{0}}$
\bea
\bar{E} = \bra{\Psi} \hat H_q \ket{\Psi} = \sum_n \bra{\Psi} \hat H_n \ket{\Psi}. 
\eea
By introducing resolutions of the identity $\hat U_n^\dagger\hat U_n$ we can rewrite $\bar{E}$ as
\bea
\bar{E} &=& \sum_n \bra{\Psi} \hat U_n^\dagger\hat U_n \hat H_n \hat U_n^\dagger\hat U_n \ket{\Psi} \\ 
&=& \sum_n \bra{\Psi} \hat U_n^\dagger \hat A_n \hat U_n \ket{\Psi} \\
&=& \sum_n \bra{\Phi_n} \hat A_n \ket{\Phi_n}  = \sum_n A_n,
\eea
where we introduced the new wavefunctions $\ket{\Phi_n} = \hat U_n \ket{\Psi}$ for which the QWC measurement 
of the $\hat A_n$ group can be done in a regular manner. Since qubit-wise commutativity always implies 
full commutativity, introducing $\hat U_n$ does not change the commutativity property of the $\hat H_n$ set. 
Therefore, if we define $\hat U_n$ and obtain $\hat A_n$ groups, we can do QWC measurements of 
$\ket{\Phi_n}$ wavefunctions that produce the expectation value of energy (see Fig.~\ref{fig:scheme}). 

To accomplish this we use a
somewhat extended idea of qubit tapering proposed by Bravyi {\it et al.}\cite{Bravyi:2017wb} for the $\hat H_n$
fragments. Bravyi {\it et al.} suggested a construction of unitary transformations applied to the whole qubit 
Hamiltonian that makes some of the qubit operators transform to the single Pauli operator $\hat x_i$, which  
allows one to substitute $\hat x_i$'s by numbers and thus to remove them from consideration. 
%(of course any other fixed single-qubit Pauli operator would work instead of $\hat x_i$). 
For our purpose, we need to substitute all single-qubit operators within $\hat H_n$ to a fixed single Pauli operator. 
This makes the transformed version of $\hat H_n$ to have only QWC terms.
Appendix B details the construction procedure for $\hat U_n$ and also proves that such transformations 
always exist for linear combinations of commuting Pauli products. Another important aspect illustrated in 
Appendix B is efficiency of implementation of $\hat U_n$ on both quantum and classical computers (see Fig.~\ref{fig:scheme}). This efficiency is guaranteed by the Gottesman-Knill theorem because $\hat U_n$'s
can be expressed as a sequence of Clifford gates.\cite{Nielsen:2010}

\subsection{Illustrative Example}

To illustrate the advantage of the new approach let us consider the model Hamiltonian
$\hat H_m = a \hat x_1\hat x_2+b\hat z_1 \hat z_2$, clearly its parts commute and share eigenstates 
(i.e. Bell states)
\bea
\ket{\parallel_{\pm}} &=& \frac{1}{\sqrt{2}} \left(\ket{\uparrow\uparrow}\pm\ket{\downarrow\downarrow}\right) \\
\ket{\perp_{\pm}} &=& \frac{1}{\sqrt{2}} \left(\ket{\uparrow\downarrow}\pm\ket{\downarrow\uparrow}\right).
\eea
These eigenstates give 
\bea
\hat z_1 \hat z_2 \ket{\parallel_{\pm}} &=& (+1) \ket{\parallel_{\pm}} \quad \hat x_1 \hat x_2 \ket{\parallel_{\pm}} = (\pm 1) \ket{\parallel_{\pm}} \\
\hat z_1 \hat z_2 \ket{\perp_{\pm}} &=&  (-1) \ket{\perp_{\pm}}\quad  \hat x_1 \hat x_2  \ket{\perp_{\pm}} = (\pm 1) \ket{\perp_{\pm}}, 
\eea
and hence their eigenvalues for the model Hamiltonians are 
\bea
\bra{\parallel_{\pm}} \hat H_m \ket{\parallel_{\pm}} &=& \pm a+ b,\\
\bra{\perp_{\pm}} \hat H_m \ket{\perp_{\pm}} &=& \pm a - b.
\eea

If $\ket{\parallel_{+}}$ is set as a trial VQE wavefunction, a single-qubit measurement 
scheme cannot determine expectation values of $\hat z_1 \hat z_2$ and $\hat x_1\hat x_2$ 
at the same time. This is easy to illustrate,
by considering $\hat z_1 \hat z_2$ measurements, as it will collapse $\ket{\parallel_{+}}$ to either 
$\ket{\uparrow\uparrow}$ or $\ket{\downarrow\downarrow}$ with equal probabilities and an eigenvalue $+1$ for 
both outcomes. However, based on only $\hat z_1 \hat z_2$ measurements we will not be able to separate 
$\ket{\parallel_{+}}$ from $\ket{\parallel_{-}}$. On the other hand, expectation values of $\hat x_1\hat x_2$ are 
uncertain after the $\hat z_1 \hat z_2$ measurement. Even though 
\bea
\bra{\uparrow\uparrow} \hat x_1\hat x_2 \ket{\uparrow\uparrow} =
\bra{\downarrow\downarrow} \hat x_1\hat x_2 \ket{\downarrow\downarrow} = 0,
\eea
there are nonzero variances for both wavefunctions. 
This is a result of a single-qubit projective measurement of $\ket{\parallel_{+}}$
that destroys the superposition and projects $\ket{\parallel_{+}}$ onto the 
$\ket{\uparrow\uparrow}$ and $\ket{\downarrow\downarrow}$ basis. 
The only way to obtain information on $\hat x_1\hat x_2 $ in the conventional single-qubit scheme 
is to start over and to measure $\ket{\parallel_{+}}$ using $\hat x_1$ and $\hat x_2$ operators. This will produce the 
second set of data because measurement of $\hat x_1$ and $\hat x_2$ operators projects a wavefunction 
to a different basis 
\bea
\ket{\parallel_{+}} &=&  \frac{1}{\sqrt{2}} \left(\ket{\rightarrow\rightarrow}+\ket{\leftarrow\leftarrow}\right), %\\
%\ket{\perp_{+}} &=&  \frac{1}{\sqrt{2}} \left(\ket{\rightarrow\rightarrow}-\ket{\leftarrow\leftarrow}\right) \\
%\ket{\parallel_{-}} &=&  \frac{1}{\sqrt{2}} \left(\ket{\leftarrow\rightarrow}+\ket{\leftarrow\rightarrow}\right) \\
%\ket{\perp_{-}} &=&  \frac{1}{\sqrt{2}} \left(\ket{\leftarrow\rightarrow}-\ket{\leftarrow\rightarrow}\right),
\eea
where $\hat x \ket{\rightarrow} = +1\ket{\rightarrow}$ and $\hat x \ket{\leftarrow} = -1\ket{\leftarrow}$.
In the case of $\ket{\parallel_{+}}$ we will obtain two projections $\ket{\rightarrow\rightarrow}$ 
and $\ket{\leftarrow\leftarrow}$ both with eigenvalue $+1$, as previously the $z$ projections 
are uncertain after measuring $\hat x_1\hat x_2 $.

The new technique introduces a unitary transformation 
$\hat U = (\hat x_1\hat x_2 + \hat z_1) (\hat z_1\hat z_2+ \hat x_2)/2$
that modifies the model Hamiltonian into a QWC group
\bea
\hat U \hat H_m \hat U^\dagger = a  \hat z_1  + b \hat x_2
\eea 
Therefore, if we measure 
\bea
\hat U\ket{\parallel_{+}} &=& \frac{1}{\sqrt{2}} \left(\ket{\uparrow\uparrow}+\ket{\uparrow\downarrow}\right) = \ket{\uparrow\rightarrow}
\eea
on the QWC operator $a  \hat z_1  + b \hat x_2$ 
we obtain $a+b$ in all instances with the wavefunction readout corresponding to $\ket{\uparrow\rightarrow}$. 
Similarly, if $\hat U\ket{\parallel_{-}} = \ket{\downarrow\rightarrow}$ is measured for the QWC operator, 
we obtain the correct answer $-a+b$ in all cases with a single set of measurements.

\subsection{Optimal partitioning of the Hamiltonian}

\begin{figure}[h!]
  \centering %
  \includegraphics[width=0.9\columnwidth]{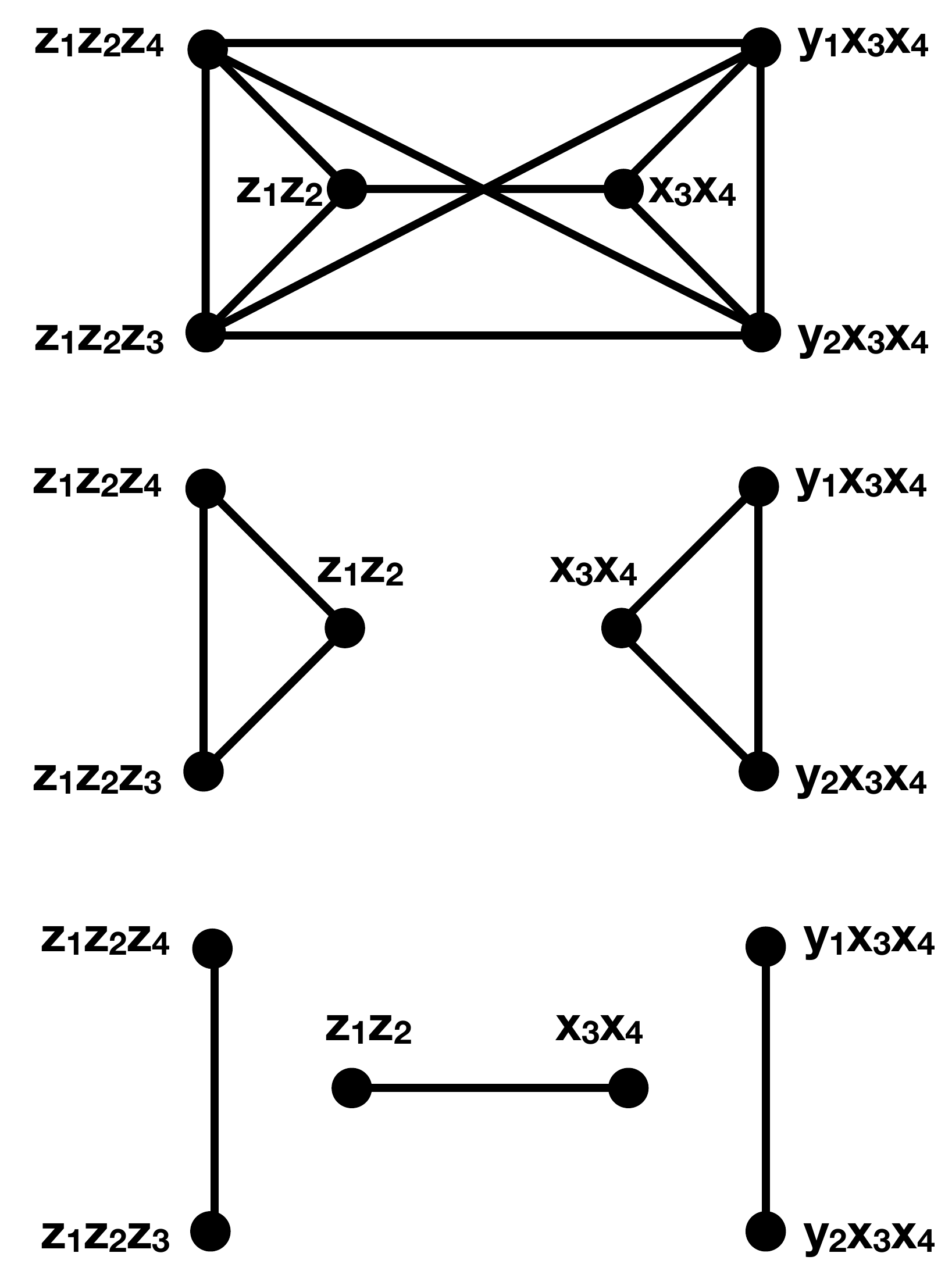}
  \caption{Graph representation of commuting terms in the Hamiltonian \eq{eq:Hex} (upper panel), minimum clique 
  cover of the graph (middle panel), non-minimum clique cover of the graph (lower panel).}
  \label{fig:FCG}
\end{figure}

Optimal partitioning for the qubit Hamiltonian to a minimal
number of groups containing mutually commuting terms can be done exactly the same way 
as in the QWC partitioning.\cite{VVpaper1} 
Regular commutativity can be also considered as a binary symmetric 
relation between Pauli products of the qubit Hamiltonian. This allows one to represent any 
qubit Hamiltonian as a graph with edges between commuting terms (vertices). 
As a simple illustration, 
one can consider the following model Hamiltonian
\bea\notag
\hat H_q &=& \hat z_1\hat z_2 + \hat z_1\hat z_2\hat z_3 + \hat z_1\hat z_2\hat z_4 \\ \label{eq:Hex}
&&+ \hat x_3\hat x_4 + \hat y_1\hat x_3\hat x_4 + \hat y_2\hat x_3\hat x_4,
\eea
whose commutativity graph is given in Fig.~\ref{fig:FCG}. Then, to determine how many terms can be measured 
at the same time, one needs to gather groups of terms that are commuting. In the graph 
representation, this means finding fully-connected sub-graphs or {\it cliques}. To optimize the measurement process 
we are interested in the minimum number of cliques, the MCC problem (Fig.~\ref{fig:FCG}, the middle panel)
\bea\notag
\hat H_q &=& \hat H_1 +\hat H_2 \\
\hat H_1 &=& \hat z_1\hat z_2 + \hat z_1\hat z_2\hat z_3 + \hat z_1\hat z_2\hat z_4 \\ 
\hat H_2 &=& \hat x_3\hat x_4 + \hat y_1\hat x_3\hat x_4 + \hat y_2\hat x_3\hat x_4.
\eea
This problem is NP-hard in general. Also, it is easy to see there are other clique covers possible 
(Fig.~\ref{fig:FCG}, the lower panel)
\bea
\hat H_q &=& \hat H_1' +\hat H_2' + \hat H_3' \\
\hat H_1' &=& \hat z_1\hat z_2 + \hat x_3\hat x_4 \\
\hat H_2' &=& \hat z_1\hat z_2\hat z_3 + \hat z_1\hat z_2\hat z_4 \\ 
\hat H_3' &=& \hat y_1\hat x_3\hat x_4 + \hat y_2\hat x_3\hat x_4.
\eea    
This solution contains larger number of cliques and thus is not optimal. 

\begin{table*}[!h]
  \caption{The number of qubits ($N$), Pauli products in the qubit Hamiltonian (Total), QWC groups ($M_{\rm QWC}$), 
  and commuting groups  produced by different heuristics (see their description in the text) for systems 
   with up to 14 qubits. The STO-3G basis has been used for all Hamiltonians unless specified otherwise. 
   BK and JW denote the Bravyi-Kitaev and Jordan-Wigner fermion-qubit transformations.}
  \label{tab:rc}
  \centering
    \begin{ruledtabular}
   \begin{tabular}{@{}lcccccccccccc@{}}
    Systems & $N$ & Total & $M_{\rm QWC}$  & GC & LF & SL & DS & RLF & DB & C & R & BKT\\
    \hline
    H$_2$ (BK)& 4 & 15 & 3 & 2 & 2  & 2 & 2 & 2 & 2 & 2 & 2 & 2\\
    LiH  (Parity)& 4 & 100 & 25 & 18 & 16 & 15 & 14 & 11 & 10 & 11 & 13 & 9\\    
    H$_2$O (6-31G, BK)& 6 & 165 & 34 &12 & 9 & 14 & 9 & 8 & 8 & 11 & 10 & 8 \\
    BeH$_2$  (BK) & 14 & 666 & 172 & 33 & 32 & 34 & 27 & 29 & 23 & 33 & 38 & -\\
BeH$_2$  (JW)& 14 & 666 & 203 & 30 & 37 & 36 & 25 & 28 & 24 & 33 & 41 & -\\
H$_2$O  (BK) & 14 & 1086 & 308 & 50 & 54 & 55 & 49 & 37 & 33 & 56 & 68 & -\\
H$_2$O (JW) & 14 & 1086 & 322 & 53 & 50 & 56 & 48 & 43 & 33 & 53 & 64 & -\\
%	N$_2$  (BK)& 20 & 2951 & 1160 & 91 & 76 & 87 & 88 & 68 & - & - & - & -\\
 %   N$_2$  (JW) & 20 & 2951 & 1187 & 90 & 81 & 95 & 90 & 76 & - & - & - & -\\
 %NH$_3$  (BK) & 16 & 3609 & 1260 & 148 & 156 & 157 & 157& 126 & - & - & - & -\\
%NH$_3$  (JW)& 16 & 3609 & 1202 & 159 & 151 & 158 & 149 & 130 & - & - & - & -\\
     \end{tabular}
       \end{ruledtabular}
\end{table*}

Even though the MCC problem is NP-hard, there are several heuristic algorithms that scale quadratically
with the number of vertices and thus can be easily used for obtaining close to optimal solutions. Assessment of 
several such heuristic techniques is done in Ref.~\citenum{VVpaper1} for Hamiltonian graphs based on qubit-wise
commutativity. Here, we will use the same heuristics as for the QWC grouping, and their descriptions can be found in Ref.~\citenum{VVpaper1} and original papers: Greedy Coloring (GC),\cite{Rigetti_doc} 
Largest First (LF),\cite{Welsh:1967} 
Smallest Last (SL),\cite{Matula:1972} DSATUR (DS),\cite{Brelaz:1973} Recursive 
Largest First (RLF),\cite{Leighton:1979} Dutton and Brigham (DB),\cite{dutton_brigham_1981} 
COSINE (C),\cite{hertz_1990}  Ramsey (R),\cite{boppana} Bron-Kerbosch-Tomita (BKT).\cite{tomita_tanaka_takahashi_2006}
All these heuristics except BKT have polynomial computational scaling with respect to the number of graph vertices.  

\begin{table*}[!h]
   \caption{Comparison of RLF results for Bravyi-Kitaev (BK) and Jordan-Wigner (JW) transformed Hamiltonians: the number of cliques ($M_C$), their maximum size (Max Size) and standard deviation of their size distribution (STD). The total number of Hamiltonian terms (Total) is almost everywhere the same for JW and BK; for the last two systems, the JW numbers are in parenthesis.}
  \label{tab:varst}
    \begin{ruledtabular}
   \begin{tabular}{@{}lcccccccc@{}}
 \multirow{2}{*}{Systems} & \multirow{2}{*}{$N$} & \multirow{2}{*}{Total} & \multicolumn{3}{c}{BK} & \multicolumn{3}{c}{JW}\\
 \cline{4-6}\cline{7-9}
	 & & & $M_C$ & Max Size & STD & $M_C$ &Max Size& STD \\
    \hline
     BeH$_2$ / STO-3G & 14 & 666 & 29 & 59 & 12.5 & 28 & 62 & 15.3\\
	H$_2$O / STO-3G & 14 & 1086 & 37 & 88 & 18.4 & 43 & 88 & 16.6\\
	NH$_3$ / STO-3G & 16 & 3609 & 126 & 92 & 15.3 & 130 & 98 & 15.6\\
	N$_2$ / STO-3G & 20 & 2951 & 68 & 128 & 26.0 & 76 & 128 & 25.1\\
    BeH$_2$ / 6-31G & 26 & 9204 & 168 & 264 & 39.8 & 163 & 312 & 46.8\\
H$_2$O / 6-31G & 26 & 12732 & 231 & 192 & 29.7 & 235 & 192 & 26.3\\
    NH$_3$ / 6-31G & 30 & 52758 (52806) & 917 & 280 & 30.4 & 922 & 238 & 29.7 \\
    N$_2$ / 6-31G & 36 & 34639 (34655) & 366 & 393 & 63.9 & 357 & 377 & 66.3 \\
     \end{tabular}
       \end{ruledtabular}
\end{table*}

\section{Numerical studies and discussion}
\label{sec:numer-stud-disc}

To assess the impact of grouping fully commuting terms we solve the MCC problem 
for graphs of qubit Hamiltonians constructed for several molecule/basis pairs 
(see Tables~\ref{tab:rc} and \ref{tab:varst}). Details of generating these Hamiltonians are given 
in Supplementary Information. 

According to Table~\ref{tab:rc}, the deviation between the minimum number of commuting cliques 
produced by different heuristics can reach up to 50\%. 
Out of all heuristics the best results on the first three small Hamiltonians were 
produced by BKT, but because of exponential scaling it is not applicable to Hamiltonians larger than hundred terms. 
The next best approach is DB, but already for the 14-qubit systems it spends 
two orders of magnitude longer of time than RLF and thus has not been selected to investigate larger systems. 
Therefore, as for QWC grouping, RLF remains the algorithm of choice, being optimal in terms of computational 
time and yielding about 25\% fewer cliques than the next-best heuristics.

\begin{figure}[h!]
  \centering %
  \includegraphics[width=1\columnwidth]{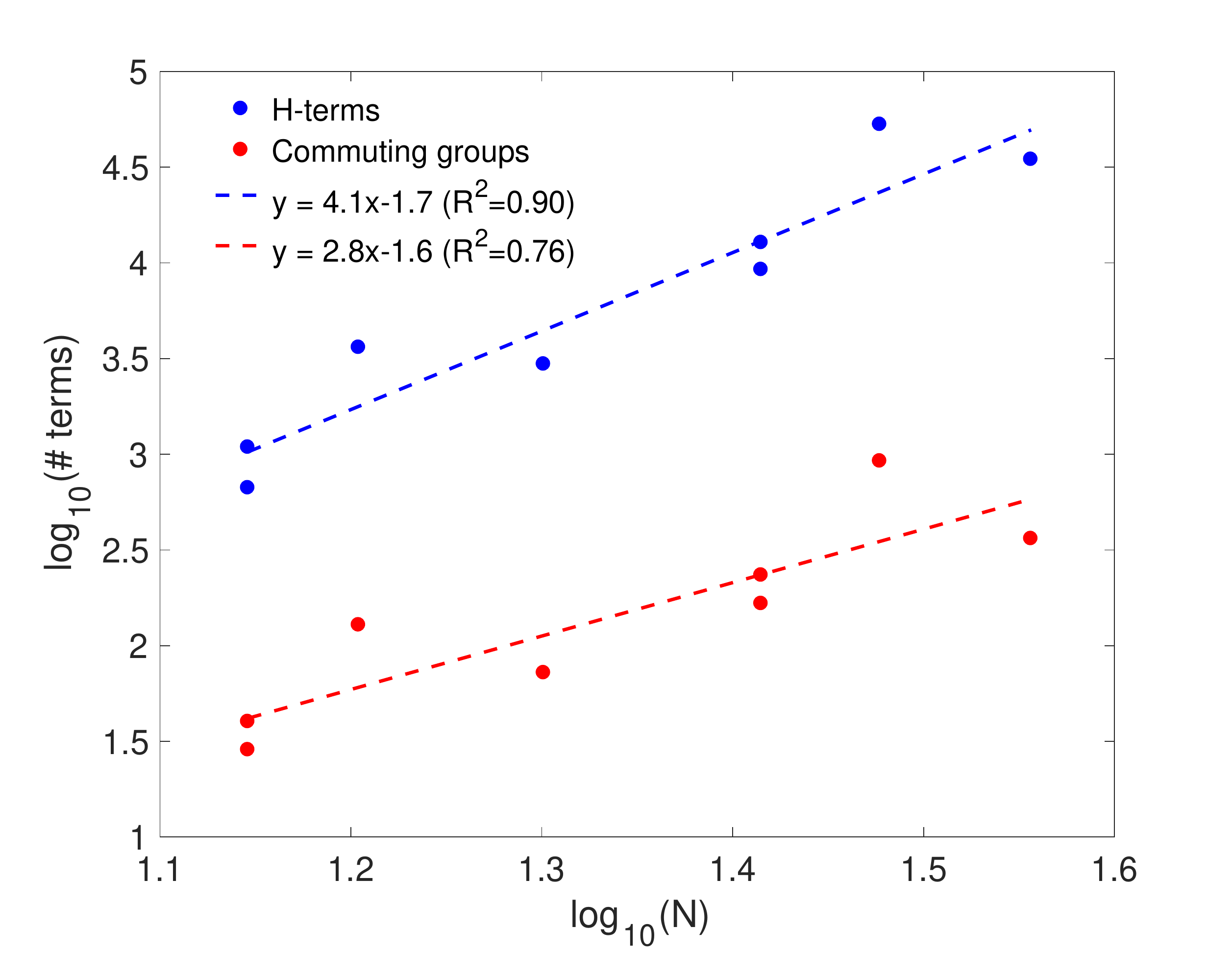}
  \caption{ Dependencies of the total number of the Hamiltonian terms (blue) 
  and the number of fully commuting groups (red) on the number of qubits for the systems in 
  Table~\ref{tab:varst} in the double log-scale. For the last two entries of Table~\ref{tab:varst}, 
  the JW and BK results were averaged.}
  \label{fig:scal}
\end{figure}

Least-square fitting of the fully commuting clique numbers for RLF in Table~\ref{tab:varst} 
with $N$ in the double logarithmic scale results in an $N^3$ dependence (\fig{fig:scal}). 
This is an $N$-fold reduction from the same dependence of the total number of terms in the studied Hamiltonians, $N^4$ (\fig{fig:scal}).
%The scaling reduction achieved by grouping all commuting terms is similar to that in our recent scheme based on grouping 
%of the anti-commuting terms\cite{} but the prefactor is three times lower in the current scheme.   
As for maximum clique sizes and standard deviations of clique size distributions (Table~\ref{tab:varst}), 
they grow with rates slightly higher and lower than linear in $N$, respectively.   
Separate analysis for the JW and BK transformations did not reveal any significant differences between 
groupings in Hamiltonians obtained with these transformations.

\section{Conclusions}
\label{sec:conclusions}

We have introduced a new method to reduce the number of measurements in the VQE approach 
 to the electronic structure problem.
The method is based on partitioning of the qubit Hamiltonian to the minimum number 
of groups whose terms are fully mutually commuting. By introducing additional 
unitary transformations each group can be transformed into a group of QWC 
terms that can be measured simultaneously. 

The main advantage of the new technique is that it 
can reduce the number of simultaneously measurable terms to largest groups of compatible  
operators without the need for modification of the currently used measuring hardware. For the considered 
examples of molecular electronic Hamiltonians the method produces at least $N$-fold reduction of the 
number of measurable groups compared to the previously used QWC grouping. 
An additional unitary transformation for each group introduces into a VQE circuit 
only $2N$ one-qubit gates and $N$ multi-qubit 
Pauli product exponents that can be decomposed into a product of up to $\sim N^{1+\log_2 3}$ two-qubit gates.
Also, since these unitary transformations modify each Pauli product into another Pauli product they can be 
encoded using only Clifford gates on a quantum computer and performed efficiently on a classical computer
according to the Gottesman-Knill theorem.\cite{Nielsen:2010} Using their Clifford property, according to Ref.~\citenum{Gottesman:CG}, the 
number of CNOT gates in circuits of the unitary transformations can be reduced to $O(N^2/log(N))$. 
In case when extra unitary transformations are not affordable due to limitations in circuit 
depth, the conventional method of grouping based on the QWC 
property\cite{Kandala:2017/nature/242,VVpaper1} is preferable, since it does not introduce 
any nonlocal gates.

Another possible application of the proposed technique is a systematic way to make nonlocal measurements 
required in mutually unbiased bases (MUB) quantum state tomography (QST).\cite{MUB_QST1,MUB_QST2}
Generally, there is an exponential growth of measurements needed in QST, a naive approach would 
require $4^{N_q}-1$ measurements, while using QWC grouping would allow for the reduction to 
$3^{N_q}$.\cite{DJ_2004} Introducing MUB is equivalent to considering fully commuting sets of Pauli 
strings, and requires only measurements of $2^{N_q}+1$ groups containing $2^{N_q}-1$ fully commuting 
Pauli strings.\cite{Zeilenger:PRA/2002} However, it is recognized that MUB-QST is challenging due to 
the entanglement present in MUB, and our unitary transformations allow one to present each of 
the $2^{N_q}+1$ groups as a QWC group and thus perform only local measurements.          

The idea of introducing unitary transformations that change some Hamiltonian fragments to the sum of 
QWC terms without modification of the expectation value can be taken to the limit 
where the whole Hamiltonian is transformed by a single unitary operator to a QWC group,
$\hat U \hat H \hat U^\dagger = \hat A$. The maximum size of a QWC group is $2^N$, and thus 
such $\hat A$ exists. This would allow one to measure the entire Hamiltonian
in a single set of measurements; however, the complexity of $\hat U$ is equivalent to that of the 
original many-body problem. Yet, this example suggests that in the measurement process one 
is not limited by only groups of fully commuting terms and more general unitary transformations 
can be potentially devised to reduce the number of simultaneously measurable terms.   
  
%\section*{Conflicts of Interest}
%There are no conflicts of interest to declare. 

%\begin{acknowledgement}
\section*{Acknowledgements}
A.F.I. acknowledges financial support from Zapata Computing Inc., the Natural Sciences and
Engineering Research Council of Canada, the Google Quantum Research Program, 
and the Mitacs Globalink Program. 
%\end{acknowledgement}

\section*{Note Added:} After submission of this manuscript to arXiv we became aware of 
several new proposals addressing the measurement problem, which appeared within a 
week or two from each other.\cite{MoscaA,IzmaylovA,BabbushA,ChicagoA,Zhao:2019vz,
BonetMonroig:2019wv,Crawford:2019tg} 

\section*{Appendix A: Commuting and Qubit-wise Commuting Terms}

Here, we illustrate that for an average Pauli product, in the full set of $4^N$ Pauli products,
there is exponentially more commuting than qubit-wise commuting Pauli products. 
An average Pauli product can be thought as $\hat e_1,..\hat e_{N/4}, \hat \sigma_{N/4+1},...\hat \sigma_{N}$, 
where $\hat e_i$'s denote the identities on $i^{\rm th}$ qubits, $\hat \sigma_j$ are Pauli operators,  
and $N \mod 4=0$ for convenience. The number of QWC terms is 
\bea
N_{\rm QWC} = 4^{N/4} 2^{3N/4} = 2^{5N/4},
\eea 
because in $\hat e_1,...\hat e_{N/4}$ one can substitute any identity by any Pauli operator without violating the 
QWC condition, which gives factor $4^{N/4}$, and in $\hat \sigma_{N/4+1},...\hat \sigma_{N}$ one can 
substitute any $\hat \sigma$ by identity, which gives rise to factor $2^{3N/4}$ . 
The number of terms commuting with the average Pauli product is
\bea
N_{\rm C}  =  4^{N}/2 = 2^{2N-1}.
\eea
Here we used the fact that the total number of Pauli products is $4^N$ 
and a half of them will be commuting with any 
non-constant Pauli product and another half will be anti-commuting. Therefore, the ratio 
$N_{\rm C}/N_{\rm QWC} = 2^{3N/4-1}$ grows exponentially with $N$.

\section*{Appendix B: Details of Unitary Transformations}

Here we detail the construction of unitary transformations that produce QWC terms from any 
linear combination of fully commuting Pauli products. A few elements of this construction are taken 
from Bravyi {\it et al.}\cite{Bravyi:2017wb}, but to keep the discussion uniform and self-contained 
we reproduce them here. 

\subsection*{Unitary transformations}

Let us consider construction of $\hat U_n$ for one of the Hamiltonian fragments $\hat H_n$ that contains 
mutually commuting Pauli products, 
\bea
\hat H_n = \sum_I C_I \hat P_I, ~[\hat P_I,\hat P_J] = 0.
\eea
Due to this commutativity and a mapping between the Pauli operator 
and symplectic linear vector spaces, it is possible to apply techniques developed by 
Bravyi {\it et al.}\cite{Bravyi:2017wb} to find a set $\mathcal{T} = \{\hat\tau_1, ..., \hat\tau_N\}$ 
of $N$ mutually commuting Pauli products, which also commute with all 
terms of $\hat H_n$. Additionally, one can find a set $\mathcal{Q} = \{\hat \sigma_1, ..., \hat \sigma_N\}$ 
of single qubit Pauli operators $\hat \sigma_i$ for each 
$\hat \tau_i$ so that 
\bea
  \{\hat \tau_i, \hat \sigma_i\} &=& 0, \\
  {[\hat \tau_i,\hat \sigma_j]} &=& 0, \quad i \neq j \label{eq:tauQComm} \\
  {[\hat \sigma_i, \hat \sigma_j]} &=& 0.
\eea 
The unitary operation $\hat U_n$ that transforms $H_n$ to its QWC form can be constructed as 
\bea
  \hat U_n &=& \prod_{i=1}^{N} \frac{1}{\sqrt{2}}(\hat\tau_i + \hat \sigma_i) \label{eq:finalUnitary}\\
  &=& \prod_{i=1}^{N} \hat V_i.
\eea 
{\BC Here, each $\hat V_i$ takes $\hat\tau_i$ to the single-qubit Pauli operator $\hat \sigma_i$ 
\bea
\hat V_i^\dagger \hat\tau_i \hat V_i &=& \frac{1}{2} (\hat\tau_i + \hat\sigma_i) \hat\tau_i (\hat\tau_i +\hat\sigma_i) \\
&=&  \frac{1}{2} (\hat\tau_i^3 +\hat\tau_i^2 \hat\sigma_i +\hat\sigma_i\hat\tau_i^2 + \hat\sigma_i\hat\tau_i \hat\sigma_i) \\
&=& \frac{1}{2} (\hat\tau_i + 2\hat\sigma_i -\hat\tau_i) = \hat\sigma_i,
\eea
where we used $\hat\tau_i^2 = \hat\sigma_i^2 = \hat e$ and anti-commutativity of $\hat\tau_i$ and $\hat\sigma_i$.
The same effect on $\hat\tau_i$ will be from the full product of $\hat V_j$'s in $\hat U_n$, 
$\hat U_n^\dagger \hat\tau_i \hat U_n = \hat\sigma_i$, because $[\hat V_j, \hat \tau_i] = 0$ if $i\ne j$.
Finally, using symplectic geometry techniques, it is possible to express every $\hat P_I$ from $\hat H_n$
as a product of $\hat \tau_i$'s up to a phase (essentially $\hat \tau_i$'s form a basis for elements of $\hat H_n$), 
and therefore, all $\hat P_I$'s can be transformed into products of $\hat \sigma_i$'s  
\bea
  \hat U_n^{\dagger}\hat H_n \hat U_n &=& \sum_I C_I \hat U_n^{\dagger} \hat P_I \hat U_n \\
  &=& \sum_I C_I \hat P_I',
\eea where $\hat P_I'$ is a product of $\hat \sigma_i \in \mathcal{Q}$ up to a phase. 
The procedures to obtain $\mathcal{T}$, $\mathcal{Q}$, and $\hat P_I$ expansions in terms of elements of 
$\mathcal{T}$ are detailed below.

\section*{Few elements of symplectic geometry}

Before going to the procedures we would like to provide few elements of symplectic geometry 
essential for understanding these procedures.}

\subsection*{Operator-vector space mapping}

To be able to use methods of symplectic geometry 
we introduce a mapping between $N$-qubit Pauli products and $2N$ symplectic vector space $\mathcal{F}$ 
over the $GF(2)$ field (also known as binary field $Z_2$).  
Any Pauli product $\hat P$ corresponds to a binary vector $\vec{P}$ with $i^{\rm th}$ and 
$(N+i)^{\rm th}$ components defined as 
\bea
  (\vec{P}_i, \vec{P}_{N+i}) = \begin{cases}
    (0, 1) & \text{$i^{\rm th}$ qubit of $\hat{P}$ is $\hat z$} \\
    (1, 0) & \text{$i^{\rm th}$ qubit of $\hat{P}$ is $\hat x$} \\ 
    (1, 1) & \text{$i^{\rm th}$ qubit of $\hat{P}$ is $\hat y$} \\
    (0, 0) & \text{$i^{\rm th}$ qubit of $\hat{P}$ is $\hat e$}. \\
  \end{cases}
\eea 
For example, for $N=4$, $\hat P = \hat x_1 \hat y_2 \hat z_3 \hat e_4$ is mapped to 
$\vec{P} = (1 1 0 0 ; 0 1 1 0)$, where the semicolon is put only for readability. 

This mapping is convenient because commutativity between two Pauli products $[\hat P_1,\hat P_2] = 0$
corresponds to orthogonality between corresponding vectors, $(\vec{P}_1|\vec{P}_2)=0$. The inner 
product $(\vec{P}_1|\vec{P}_2)$ between two vectors $\vec{P}_1$ and $\vec{P}_2$ 
is defined in a symplectic manner 
\bea
(\vec{P}_1|\vec{P}_2) = ( \vec{P}_1, \mathbf{J}  \vec{P}_2), 
\eea
where $(.,.)$ is the normal Euclidian inner product and $\mathbf{J}$ is a symplectic metric matrix
\bea
\mathbf{J} = \begin{bmatrix}
        \mathbf{0}_{N\times N}, \mathbf{1}_{N\times N} \\
        \mathbf{1}_{N\times N}, \mathbf{0}_{N\times N}
    \end{bmatrix}.
\eea
Therefore, we will use commutation and orthogonality interchangeably. The symplectic 
inner product is bi-linear, and thus if $\hat P_1$ commutes with $\hat P_2$ and $\hat P_3$, 
then $\vec{P}_1$ is orthogonal to $\vec{P}_2 + \vec{P}_3$.
Also anti-commutation $\{\hat P_1,\hat P_2\} = 0$ corresponds to $(\vec{P}_1|\vec{P}_2) = 1$.

Another useful correspondence is between results of addition of vectors and multiplication of Pauli product operators: 
$\vec{P}_1 + \vec{P}_2 = \vec{P}_3$ is equivalent to $\hat P_1\cdot\hat P_2 = p \hat P_3$, 
where $p$ is a phase factor that has values $\pm 1, \pm i$ depending on single-qubit Pauli operators and their order 
in the $\hat P_1\hat P_2$ product.  

\subsection*{Types of symplectic subspaces}
Here we introduce a few types of symplectic subspaces that will be utilized to treat the fully commuting
sets of operators. 
For a subspace $V$, the orthogonal complement will be denoted by $V^{\perp}$. 
The dimensionalities of the two subspaces are connected by $dim(V) + dim(V^{\perp}) = 2N$,
while taking the complement twice returns the initial subspace, $(V^{\perp})^{\perp} = V$.
The examples below are given for $N=2$, which corresponds to a 4-dimensional symplectic vector space. 
\begin{itemize}
    % Example
    \item $V$ is Isotropic $\leftrightarrow V \subset V^{\perp}$. For example, $V = span\{(10 ; 00)\}$
    is an isotropic subspace with the orthogonal complement  
    \bea
        V^{\perp} &=& span\{(10; 00), (01; 00), (00; 01)\},
    \eea which contains $V$. 
    % Example
    \item $V$ is Coisotropic $\leftrightarrow V^{\perp} \subset V$. For example, taking the 
    $V^\perp$ of the previous example as $V$ one obtains a coisotropic subspace 
    \bea
        V &=& span\{(10; 00), (01; 00), (00; 01)\}, \\
        V^{\perp} &=& span\{(10 ; 00)\}.
    \eea
    \item $V$ is Lagrangian $\leftrightarrow V = V^{\perp}$. For example,
    \bea
        V &=& span\{(10; 00), (01; 00)\} = V^{\perp}.
    \eea 
    A Lagrangian subspace is also the largest isotropic or the smallest coisotropic subspace. 
    % Example
\end{itemize}
Based on these examples it is clear that 
if $V$ is isotropic then $V^{\perp}$ is coisotropic, and $dim(V)\leq N$, $dim(V^{\perp})\geq N$. 
It also can be proven that, for any isotropic space $V$, there exists Lagrangian subspace $L$ such that 
$V \subseteq L \subseteq V^{\perp}$, and $dim(L) = N$.\cite{SympGeomBook}

\section*{Procedures}

%\section*{Appendix E: Existence}
%\subsection{Application to fully commuting subsets}

A set of mutually commuting Pauli products from $\hat H_n$ is isomorphic to  
an isotropic subspace in the symplectic vector space of $2N$ dimensions. 
Thus, we can always find the Lagrangian subspace and its basis of $N$ orthogonal basis vectors. 
These basis vectors are mapped to the mutually commuting $\hat \tau_i$'s operators 
that also commute with all terms in $\hat H_n$.

\subsection*{Finding $\hat \tau_i$'s}

Here we show how to find the $N$ mutually commuting operators $\hat \tau_i$'s that also commute
with all terms in the group of Pauli terms $\hat H_n$. 

% Is mutual commutation a sufficient condition for isotropic subspace? 
% Simple. Check if each term mutually commute, 
% Then their basis must also mutually commute. Obvious in vector representation.
{\it Step 1: Finding orthogonal basis vectors using Gaussian elimination:}
%Condition: Check if basis forms an isotropic subspace}
%Procedure simply finds the basis of $\hat H_n$ in $\mathcal{F}$ and check if they mutually commute.  
%This is equivalent to checking if all terms mutually commute because addition in $\mathcal{F}$ 
%does not change mutual commutation.
Gaussian elimination for all elements of $\hat H_n$ creates a basis of $V$ that is mutually commuting because 
the original terms are mutually commuting and their addition does not change this property. 
This basis forms isotropic space because all basis vectors are self-orthogonal and therefore 
can be thought as a part of the $V^\perp$ basis, hence, the condition $V\subseteq V^\perp$ is satisfied
and $V$ is isotropic. $dim(V)=K\le N$, and if  $K=N$ then $V$ is Lagrangian and there is no need to 
do anything else, the $\hat \tau_i$'s are obtained from the $N$ basis vectors of $V$. If $K<N$ then 
the next step of building the basis of the orthogonal complement $V^\perp$ is needed.
 
% Simple null space
{\it Step 2: Finding the basis for $V^\perp$:}
The normal binary null space is obtained for $V$. Then the first and second halves of the indices 
are interchanged so that we obtain a null space in the symplectic sense. 
This null space is $V^\perp$, it is coisotropic, and $dim(V^\perp)=M\in[N+1,2N)$.
$M\neq N$ because $M=N$ would require $K=N$, and thus the procedure would have exited 
on the first step.

% Symplectic Gram-Schmidt
{\it Step 3: Finding the Lagrangian subspace in $V^\perp$:}
%If $M=N$, then $V^\perp$ is already Lagrangian, and its basis vectors are mapped directly to $\hat \tau_i$'s. 
$V^\perp$ is coisotropic and therefore some of its basis vectors commute and the others anti-commute.
To obtain $N$ mutually commuting vectors (the basis of the Lagrangian subspace), this step eliminates $M-N$ vectors 
from the basis of $V^\perp$, $\{\vec{c}_i\}_{i=1}^{M}$, by using a symplectic version of the Gram-Schmidt 
orthogonalization procedure. First, a pair of anti-commuting vectors 
is found. Using the enumeration freedom we can assume that this pair is formed 
by first two vectors: $(\vec{c}_1|\vec{c}_2)=1$.
Then the other vectors are orthogonalized to the first two as follows
\bea
    {\vec{c'}_k} = \vec{c}_k + (\vec{c}_k| \vec{c}_2)\vec{c}_1 + (\vec{c}_k| \vec{c}_1)\vec{c}_2, ~k \in[3, M]
\eea 
so that 
\bea
    (\vec{c}_1| \vec{c'}_k) = (\vec{c}_2| \vec{c'}_k) = 0.
\eea 
Then a new basis set of $M-1$ vectors is formed, $\vec{c}_1\cup\{\vec{c'}_k\}_{k=3}^M$.
Note that there is a freedom in eliminating either $\vec{c}_1$ or $\vec{c}_2$ from the old basis.  
In the new basis, the only possible source of anti-commutativity is the $\{\vec{c'}_k\}$ subset, so the 
procedure for the search of an anti-commuting pair is repeated. Once the new pair is found
 the procedure of orthogonalization of all $\{\vec{c'}_k\}$ to that pair is repeated with elimination 
 of one of the pair members to produce $M-2$ basis vectors. 
Once the algorithm cannot find any new anti-commuting pairs, it will have $N$ mutually commuting 
basis vectors of the Lagrangian subspace that can be mapped directly to $\hat \tau_i$'s. 

% Are we guranteed the existence of these single qubit terms. By construction
\subsection*{Finding $\hat \sigma_i$'s}

Given a set of $\hat \tau_i \in\mathcal{T}$, to build 
the unitary transformation $\hat U_n$ [\eq{eq:finalUnitary}] requires 
a set of single qubit Pauli operators 
$\hat \sigma$ (i.e. $x_1,\ y_2,\ z_3$ {\it etc.}) satisfying  
\bea
    (\vec{\tau}_i| \vec{\sigma}_j) &=& \begin{cases}
        1, & i = j \\
        0, & i \neq j
    \end{cases} \label{eq:TPanti}\\
    (\vec{\sigma}_i | \vec{\sigma}_j ) &=& 0. \label{eq:PDiffQub}
\eea 
Note that \eq{eq:PDiffQub} requires that all $\vec{\sigma}_i$'s correspond to different qubits. 
The resulting transformation $\hat U_n$ will transform $\hat \tau_i  \rightarrow \hat \sigma_i$.

For $\hat \tau_1$, we have $N$ qubits available to define $\hat \sigma_1$, so 
if $\hat \tau_1$ has $\hat x_1$, $\hat \sigma_1$ should be one of anti-commuting operators $\hat y_1$
or $\hat z_1$. To make the rest of $\hat \tau_i$'s to commute with $\hat \sigma_1$ we perform 
an orthogonalization step: 
\bea\label{eq:ortho}
    \vec{\tau'}_k = \vec{\tau}_k + (\vec{\tau}_k| \vec{\sigma}_1) \vec{\tau}_1,~  k \in[2,N], \label{eq:tauMod}
\eea 
so that $( \vec{\tau'}_k| \vec{\sigma_1}) = 0$ is guaranteed, and 
the mutual commutation between $\vec{\tau}_k$ is not changed.
Then we find $\hat \sigma_i$'s for the available qubits for the rest of $\hat \tau'_i$ 
and after finding each $\hat \sigma_i$ we do re-orthogonalization of $\{\hat \tau'_j\}_{j=i+1}^N$.

Let us prove the existence of $N$ $\hat \sigma_i$'s that can be found in the described process.  
We have already shown that it is straightforward to find the initial $\hat \sigma_1$, let us consider 
some intermediate step, where $\mathcal{T}_A$ is a subset of $\mathcal{T}$ with $N_A$ elements 
for which $N_A$ $\hat \sigma_i$'s are found and are assigned to a $\mathcal{Q}_A$ set. 
Then $\mathcal{T}_B = \mathcal{T}/\mathcal{T}_A$ is a complementary subset with the rest of 
$\hat \tau_i$'s that are commuting with operators from both $\mathcal{T}_A$ and $\mathcal{Q}_A$ sets. 
To continue the process we need to find a non-trivial (i.e. non-identity) qubit operator $\hat \sigma'$ for 
a qubit that is not present in the $\mathcal{Q}_A$ set but is in one of the elements of $\mathcal{T}_B$. Then, 
constructing the next $\hat \sigma_{N_A+1}$ operator 
can be done by taking an operator that anti-commutes with $\hat \sigma'$. 
To prove that this is possible we will show that such a non-trivial $\hat \sigma'$ exists.
Let us assume the contrary and arrive to a contradiction. Indeed, if all $\mathcal{T}_B$ elements
have only trivial (i.e. identity) qubit operators for qubits higher than $N_A$, then they either must be all
equal to the identity, which is a zero vector and cannot be a basis vector in the Lagrangian subspace, 
or they will not be able to commute with both $\mathcal{Q}_A$ and $\mathcal{T}_A$ sets simultaneously, 
which is also a contradiction to the initial assumption about commutativity of $\mathcal{T}_B$ elements 
with the $\mathcal{Q}_A$ and $\mathcal{T}_A$ sets. 

{\BC
\subsection*{Obtaining expansions in $\hat \tau_i$'s}

To apply the unitary operator $\hat U_n$ in \eq{eq:finalUnitary} to all Pauli products $\hat P_I$'s in $\hat H_n$, 
$\hat P_I$'s need to be presented as products of $\hat \tau_i$'s. 
Let $\hat P$ be one of the Pauli products in $\hat H_n$ for which sets $\mathcal{T}$  
and $\mathcal{Q}$ are found. Since the $\mathcal{T}$-set forms a basis in $L$ and the sum of the vectors can be 
mapped to a product of operators up to a phase, we can always find a subset, $\mathcal{K} \subset \mathcal{T}$, so that 
\bea\label{eq:ph}
  \hat P = p \prod_{\hat\tau_k \in \mathcal{K}} \hat\tau_k,
\eea 
where $p$ is a phase that arises from multiplication between $\hat \tau_k$. 
To find subset $\mathcal{K}$ we solve a system of linear equations $\mathbf{A}\bar{x} = \bar{b}$,
where $\mathbf{A} = [\tau_1,...,\tau_N]$ is the $2N\times N$ binary matrix built out of $\hat \tau_i$'s operators turned into
corresponding vectors, and $\bar{b}$ is a symplectic vector representation of $\hat P$. Indices of the binary vector
$\bar{x}$ corresponding to nonzero entries give $k$ indices for $\hat\tau_k$'s in $\mathcal{K}$. We use the 
Gaussian elimination to solve for $\bar{x}$ in binary field $GF(2)$.
This representation makes transformation of each $\hat P$ analytic
\bea
  \hat U_n^{\dagger} \hat P \hat U_n &=& p \prod_{\hat\tau_k \in \mathcal{K}} \hat U_n^{\dagger} \hat\tau_k \hat U_n \\
  &=& p \prod_k \hat \sigma_k. \label{eq:polyShortCut}
\eea 

\subsection*{Complexity of operations performed on a classical computer}

Finding the partitioning of the Hamiltonian into groups of all commuting terms uses the RLF 
heuristic that scales cubicly with the number of terms in $\hat H$, the total number of terms in 
the Hamiltonian scales as $O(N^4)$, and therefore, the total scaling of this step is $O(N^{12})$.
In spite of the large degree, the pre-factor of this scaling is quite small.%, $\sim 10^{-9}$.

Here we will consider the computational cost for finding $\hat U_n$ for each of fully-commuting 
groups of terms in the Hamiltonian, $\hat H_n$. Let $n$ be the number of $\hat P_I$ terms in $\hat H_n$.  
Constructing $\hat U_n$ requires the following procedures:

1) Finding $N$ $\hat\tau_i$'s: The preliminary step 
involves representing $n$ $\hat P_I$'s in the binary form, which scales as $O(nN)$.

{\it Step 1:} The Gaussian elimination applied to the binary matrix of $\vec{P}_I$'s 
(the matrix size is $n\times 2N$) costs $O(nN^2 )$.

{\it Step 2:} Obtaining the null space after the Gaussian elimination costs $O(N^2)$.

{\it Step 3:} Constructing the Lagrangian subspace costs $O(N^3)$ in CPU and $O(N^2)$ in memory
to save intermediate results of anti-commuting pairs. 

2) Finding $\hat \sigma_i$'s: The orthogonalization step in \eq{eq:ortho} leads to $O(N^3)$ scaling of $\hat \sigma_i$'s search
because for each of $N$ $\hat\tau_i$'s, a particular $\hat\sigma_i$ can lead to modification of at most $N$ 
other $\hat \tau_i$'s and each modification scales as $O(N)$. 

3) Expanding $\hat P_I$’s in $\hat \tau_i$'s: Finding $P_i$ as a product of $\hat \tau_k$ 
uses the Gaussian elimination, which costs $O(N^3)$ for each element; there are 
$n$ elements, so the total cost is $O(nN^3)$.  
Computing the phase $p$ in \eq{eq:ph} for each $\hat P_I$ requires $O(N^2)$ multiplications 
since the number of $\hat \tau_k$'s scales as $O(N)$ and each of them can have up to $N$ 
Pauli operators. Thus, this part scales as $O(nN^2)$ in total. 

Considering all parts of the $\hat U_n$ generation, the largest scaling is $O(nN^3)$.

}

\subsection*{Encoding unitary transformations on a quantum computer}

To put $\hat V_i$'s into a form acceptable for encoding on a quantum computer, 
we rewrite them as 
\bea
    \hat V_i &=& (-i)e^{i \frac{\pi}{4} \hat\sigma_i}e^{i \frac{\pi}{4}\hat \tau_i}e^{i \frac{\pi}{4}\hat\sigma_i}.
\eea 
It is straightforward to check that it is indeed an equality 
    \bea
    e^{i \frac{\pi}{4} \hat\sigma_i}e^{i \frac{\pi}{4}\hat \tau_i}e^{i \frac{\pi}{4}\hat\sigma_i} &=& \frac{1}{2^{3/2}} ({1+i\hat\sigma_i})({1+i\hat \tau_i})({1+i\hat\sigma_i}) \notag \\
    &=& \frac{1}{2^{3/2}} ({1 + i\hat\sigma_i + i\hat\tau_i - \hat\sigma_i\hat\tau_i})({1 + i\hat\sigma_i}) \notag \\
    &=& \frac{1}{2^{3/2}} (1 + i\hat\sigma_i + i\hat\tau_i \notag \\
    &&- \hat\sigma_i\hat\tau_i + i\hat\sigma_i - 1 - \hat\tau_i\hat\sigma_i - i\hat\sigma_i\hat\tau_i\hat\sigma_i) \notag \\
    &=& \frac{i}{\sqrt{2}} ({\hat\tau_i + \hat \sigma_i}) = i \hat V_i,
\eea 
where to arrive at the last line we used anti-commutation between $\hat\tau_i$ and $\hat \sigma_i$.

In the worst case $\hat\tau_i$ may involve all $N$ qubits. We can decompose $e^{i \frac{\pi}{4} \hat\tau_i}$ 
into product of $O(N^{\log_2 3})$ two-qubit operations\cite{Ryabinkin:2018/qcc}. 
Hence, $\hat U_n$ [\eq{eq:finalUnitary}] requires at most $O(N^{1 + \log_2 3})$ one- and two-qubit gates.
Moreover, note that $\hat U_n$ transforms each Pauli product into another Pauli product and thus 
can be written as a sequence of Clifford gates.\cite{Nielsen:2010} 
Aaronson and Gottesman\cite{Gottesman:CG} found that for circuits containing only Clifford gates the   
number of CNOT gates can be bounded by $O(N^2/log(N))$.

\subsection*{Example: H$_2$ molecule}

\label{sec:H2_results}
We provide below a simple example of constructing the unitary transformation for 
one of two mutually commuting groups in the qubit Hamiltonian of \ce{H2}/STO-3G.
The BK transformed qubit Hamiltonian of \ce{H2} contains the following mutually commuting group  
\bea\notag
\hat H_{1} &=& -0.4738 + 0.1412z_1 + 0.0558x_2z_1x_0 + 0.0558y_2z_1y_0 \\ 
&& + 0.0868z_2z_0 + 0.1425z_2z_1z_0 + 0.1489z_3z_1 \notag \\
&& + 0.0558z_3x_2z_1x_0 + 0.0558z_3y_2z_1y_0 + 0.0868z_3z_2z_0 \notag \\
&& + 0.1425z_3z_2z_1z_0.
\eea
The described procedures produce the following sets of $\mathcal{T}$ and $\mathcal{Q}$ 
\bea
 \mathcal{T} &=& \{\hat z_3,\ \hat z_1,\ \hat y_2\hat y_0,\ \hat x_2\hat x_0\} \\
\mathcal{Q} &=& \{\hat x_3,\ \hat x_1,\ \ \hat x_2,\quad \hat y_0\},
\eea 
and the unitary operation 
\bea
 \hat U_1 &=& 0.25x_3x_1x_0 + 0.25x_3z_1x_0 + 0.25x_3x_2x_1y_0 \notag\\ 
  &&+ 0.25x_3x_2z_1y_0 + 0.25x_3y_2x_1 + 0.25x_3y_2z_1 \notag\\
  &&- 0.25x_3z_2x_1z_0 - 0.25x_3z_2z_1z_0 + 0.25z_3x_1x_0 \notag\\
  &&+ 0.25z_3z_1x_0 + 0.25z_3x_2x_1y_0 + 0.25z_3x_2z_1y_0 \notag\\
  &&+ 0.25z_3y_2x_1 + 0.25z_3y_2z_1 - 0.25z_3z_2x_1z_0 \notag\\
  &&- 0.25z_3z_2z_1z_0.
\eea 
The result of the transformation is a QWC group 
\bea
  \hat U^{\dagger} H_1 \hat U &=& -0.4738 + 0.1412x_1 + 0.0558x_1y_0 \notag\\ 
  && - 0.0868x_2y_0 + 0.0558x_2x_1 - 0.1425x_2x_1y_0 \notag\\ 
  &&+ 0.1489x_3x_1 + 0.0558x_3x_1y_0 - 0.0868x_3x_2y_0 \notag\\ 
  &&+ 0.0558x_3x_2x_1 - 0.1425x_3x_2x_1y_0.
\eea 
%which agrees with the results of the more efficient procedure in \eq{eq:polyShortCut}.
%\subsection{LiH molecule}
%\label{sec:LiH_results}
%\subsection{H2O molecule}
%\label{sec:H2O_results}

%\bibliography{qcomp-snap,books,programs,databases,addVV}

\begin{thebibliography}{49}%
\makeatletter
\providecommand \@ifxundefined [1]{%
 \@ifx{#1\undefined}
}%
\providecommand \@ifnum [1]{%
 \ifnum #1\expandafter \@firstoftwo
 \else \expandafter \@secondoftwo
 \fi
}%
\providecommand \@ifx [1]{%
 \ifx #1\expandafter \@firstoftwo
 \else \expandafter \@secondoftwo
 \fi
}%
\providecommand \natexlab [1]{#1}%
\providecommand \enquote  [1]{``#1''}%
\providecommand \bibnamefont  [1]{#1}%
\providecommand \bibfnamefont [1]{#1}%
\providecommand \citenamefont [1]{#1}%
\providecommand \href@noop [0]{\@secondoftwo}%
\providecommand \href [0]{\begingroup \@sanitize@url \@href}%
\providecommand \@href[1]{\@@startlink{#1}\@@href}%
\providecommand \@@href[1]{\endgroup#1\@@endlink}%
\providecommand \@sanitize@url [0]{\catcode `\\12\catcode `\$12\catcode
  `\&12\catcode `\#12\catcode `\^12\catcode `\_12\catcode `\%12\relax}%
\providecommand \@@startlink[1]{}%
\providecommand \@@endlink[0]{}%
\providecommand \url  [0]{\begingroup\@sanitize@url \@url }%
\providecommand \@url [1]{\endgroup\@href {#1}{\urlprefix }}%
\providecommand \urlprefix  [0]{URL }%
\providecommand \Eprint [0]{\href }%
\providecommand \doibase [0]{http://dx.doi.org/}%
\providecommand \selectlanguage [0]{\@gobble}%
\providecommand \bibinfo  [0]{\@secondoftwo}%
\providecommand \bibfield  [0]{\@secondoftwo}%
\providecommand \translation [1]{[#1]}%
\providecommand \BibitemOpen [0]{}%
\providecommand \bibitemStop [0]{}%
\providecommand \bibitemNoStop [0]{.\EOS\space}%
\providecommand \EOS [0]{\spacefactor3000\relax}%
\providecommand \BibitemShut  [1]{\csname bibitem#1\endcsname}%
\let\auto@bib@innerbib\@empty
%</preamble>
\bibitem [{\citenamefont {Preskill}(2018)}]{Preskill:2018jf}%
  \BibitemOpen
  \bibfield  {author} {\bibinfo {author} {\bibfnamefont {J.}~\bibnamefont
  {Preskill}},\ }\href@noop {} {\bibfield  {journal} {\bibinfo  {journal}
  {Quantum}\ }\textbf {\bibinfo {volume} {2}},\ \bibinfo {pages} {79} (\bibinfo
  {year} {2018})}\BibitemShut {NoStop}%
\bibitem [{\citenamefont {Peruzzo}\ \emph {et~al.}(2014)\citenamefont
  {Peruzzo}, \citenamefont {McClean}, \citenamefont {Shadbolt}, \citenamefont
  {Yung}, \citenamefont {Zhou}, \citenamefont {Love}, \citenamefont
  {Aspuru-Guzik},\ and\ \citenamefont {O'Brien}}]{Peruzzo:2014/ncomm/4213}%
  \BibitemOpen
  \bibfield  {author} {\bibinfo {author} {\bibfnamefont {A.}~\bibnamefont
  {Peruzzo}}, \bibinfo {author} {\bibfnamefont {J.}~\bibnamefont {McClean}},
  \bibinfo {author} {\bibfnamefont {P.}~\bibnamefont {Shadbolt}}, \bibinfo
  {author} {\bibfnamefont {M.-H.}\ \bibnamefont {Yung}}, \bibinfo {author}
  {\bibfnamefont {X.-Q.}\ \bibnamefont {Zhou}}, \bibinfo {author}
  {\bibfnamefont {P.~J.}\ \bibnamefont {Love}}, \bibinfo {author}
  {\bibfnamefont {A.}~\bibnamefont {Aspuru-Guzik}}, \ and\ \bibinfo {author}
  {\bibfnamefont {J.~L.}\ \bibnamefont {O'Brien}},\ }\href@noop {} {\bibfield
  {journal} {\bibinfo  {journal} {Nat. Commun.}\ }\textbf {\bibinfo {volume}
  {5}},\ \bibinfo {pages} {4213} (\bibinfo {year} {2014})}\BibitemShut
  {NoStop}%
\bibitem [{\citenamefont {McArdle}\ \emph {et~al.}(2018)\citenamefont
  {McArdle}, \citenamefont {Endo}, \citenamefont {Aspuru-Guzik}, \citenamefont
  {Benjamin},\ and\ \citenamefont {Yuan}}]{McArdle:2018we}%
  \BibitemOpen
  \bibfield  {author} {\bibinfo {author} {\bibfnamefont {S.}~\bibnamefont
  {McArdle}}, \bibinfo {author} {\bibfnamefont {S.}~\bibnamefont {Endo}},
  \bibinfo {author} {\bibfnamefont {A.}~\bibnamefont {Aspuru-Guzik}}, \bibinfo
  {author} {\bibfnamefont {S.}~\bibnamefont {Benjamin}}, \ and\ \bibinfo
  {author} {\bibfnamefont {X.}~\bibnamefont {Yuan}},\ }\href@noop {} {\bibfield
   {journal} {\bibinfo  {journal} {arXiv.org}\ ,\ \bibinfo {pages}
  {arXiv:1808.10402}} (\bibinfo {year} {2018})},\ \Eprint
  {http://arxiv.org/abs/1808.10402v1} {1808.10402v1} \BibitemShut {NoStop}%
\bibitem [{\citenamefont {Cao}\ \emph {et~al.}(2018)\citenamefont {Cao},
  \citenamefont {Romero}, \citenamefont {Olson}, \citenamefont {Degroote},
  \citenamefont {Johnson}, \citenamefont {Kieferov{\'a}}, \citenamefont
  {Kivlichan}, \citenamefont {Menke}, \citenamefont {Peropadre}, \citenamefont
  {Sawaya}, \citenamefont {Sim}, \citenamefont {Veis},\ and\ \citenamefont
  {Aspuru-Guzik}}]{Cao:2018vo}%
  \BibitemOpen
  \bibfield  {author} {\bibinfo {author} {\bibfnamefont {Y.}~\bibnamefont
  {Cao}}, \bibinfo {author} {\bibfnamefont {J.}~\bibnamefont {Romero}},
  \bibinfo {author} {\bibfnamefont {J.~P.}\ \bibnamefont {Olson}}, \bibinfo
  {author} {\bibfnamefont {M.}~\bibnamefont {Degroote}}, \bibinfo {author}
  {\bibfnamefont {P.~D.}\ \bibnamefont {Johnson}}, \bibinfo {author}
  {\bibfnamefont {M.}~\bibnamefont {Kieferov{\'a}}}, \bibinfo {author}
  {\bibfnamefont {I.~D.}\ \bibnamefont {Kivlichan}}, \bibinfo {author}
  {\bibfnamefont {T.}~\bibnamefont {Menke}}, \bibinfo {author} {\bibfnamefont
  {B.}~\bibnamefont {Peropadre}}, \bibinfo {author} {\bibfnamefont {N.~P.~D.}\
  \bibnamefont {Sawaya}}, \bibinfo {author} {\bibfnamefont {S.}~\bibnamefont
  {Sim}}, \bibinfo {author} {\bibfnamefont {L.}~\bibnamefont {Veis}}, \ and\
  \bibinfo {author} {\bibfnamefont {A.}~\bibnamefont {Aspuru-Guzik}},\
  }\href@noop {} {\bibfield  {journal} {\bibinfo  {journal} {arXiv.org}\ ,\
  \bibinfo {pages} {arXiv:1812.09976}} (\bibinfo {year} {2018})},\ \Eprint
  {http://arxiv.org/abs/1812.09976v2} {1812.09976v2} \BibitemShut {NoStop}%
\bibitem [{\citenamefont {Aspuru-Guzik}\ \emph {et~al.}(2005)\citenamefont
  {Aspuru-Guzik}, \citenamefont {Dutoi}, \citenamefont {Love},\ and\
  \citenamefont {Head-Gordon}}]{AspuruGuzik:2005/sci/1704}%
  \BibitemOpen
  \bibfield  {author} {\bibinfo {author} {\bibfnamefont {A.}~\bibnamefont
  {Aspuru-Guzik}}, \bibinfo {author} {\bibfnamefont {A.~D.}\ \bibnamefont
  {Dutoi}}, \bibinfo {author} {\bibfnamefont {P.~J.}\ \bibnamefont {Love}}, \
  and\ \bibinfo {author} {\bibfnamefont {M.}~\bibnamefont {Head-Gordon}},\
  }\href {\doibase 10.1126/science.1113479} {\bibfield  {journal} {\bibinfo
  {journal} {Science}\ }\textbf {\bibinfo {volume} {309}},\ \bibinfo {pages}
  {1704} (\bibinfo {year} {2005})}\BibitemShut {NoStop}%
\bibitem [{\citenamefont {Babbush}\ \emph {et~al.}(2014)\citenamefont
  {Babbush}, \citenamefont {Love},\ and\ \citenamefont
  {Aspuru-Guzik}}]{Babbush:2014/sr/6603}%
  \BibitemOpen
  \bibfield  {author} {\bibinfo {author} {\bibfnamefont {R.}~\bibnamefont
  {Babbush}}, \bibinfo {author} {\bibfnamefont {P.~J.}\ \bibnamefont {Love}}, \
  and\ \bibinfo {author} {\bibfnamefont {A.}~\bibnamefont {Aspuru-Guzik}},\
  }\href {\doibase 10.1038/srep06603} {\bibfield  {journal} {\bibinfo
  {journal} {Sci. Rep.}\ }\textbf {\bibinfo {volume} {4}},\ \bibinfo {pages}
  {6603} (\bibinfo {year} {2014})}\BibitemShut {NoStop}%
\bibitem [{\citenamefont {Kandala}\ \emph {et~al.}(2017)\citenamefont
  {Kandala}, \citenamefont {Mezzacapo}, \citenamefont {Temme}, \citenamefont
  {Takita}, \citenamefont {Brink}, \citenamefont {Chow},\ and\ \citenamefont
  {Gambetta}}]{Kandala:2017/nature/242}%
  \BibitemOpen
  \bibfield  {author} {\bibinfo {author} {\bibfnamefont {A.}~\bibnamefont
  {Kandala}}, \bibinfo {author} {\bibfnamefont {A.}~\bibnamefont {Mezzacapo}},
  \bibinfo {author} {\bibfnamefont {K.}~\bibnamefont {Temme}}, \bibinfo
  {author} {\bibfnamefont {M.}~\bibnamefont {Takita}}, \bibinfo {author}
  {\bibfnamefont {M.}~\bibnamefont {Brink}}, \bibinfo {author} {\bibfnamefont
  {J.~M.}\ \bibnamefont {Chow}}, \ and\ \bibinfo {author} {\bibfnamefont
  {J.~M.}\ \bibnamefont {Gambetta}},\ }\href {\doibase 10.1038/nature23879}
  {\bibfield  {journal} {\bibinfo  {journal} {Nature}\ }\textbf {\bibinfo
  {volume} {549}},\ \bibinfo {pages} {242} (\bibinfo {year}
  {2017})}\BibitemShut {NoStop}%
\bibitem [{\citenamefont {Genin}\ \emph {et~al.}(2019)\citenamefont {Genin},
  \citenamefont {Ryabinkin},\ and\ \citenamefont {Izmaylov}}]{Genin:2019te}%
  \BibitemOpen
  \bibfield  {author} {\bibinfo {author} {\bibfnamefont {S.~N.}\ \bibnamefont
  {Genin}}, \bibinfo {author} {\bibfnamefont {I.~G.}\ \bibnamefont
  {Ryabinkin}}, \ and\ \bibinfo {author} {\bibfnamefont {A.~F.}\ \bibnamefont
  {Izmaylov}},\ }\href@noop {} {\bibfield  {journal} {\bibinfo  {journal}
  {arXiv.org}\ ,\ \bibinfo {pages} {arXiv:1901.04715}} (\bibinfo {year}
  {2019})},\ \Eprint {http://arxiv.org/abs/1901.04715v1} {1901.04715v1}
  \BibitemShut {NoStop}%
\bibitem [{\citenamefont {Helgaker}\ \emph {et~al.}(2000)\citenamefont
  {Helgaker}, \citenamefont {Jorgensen},\ and\ \citenamefont
  {Olsen}}]{Helgaker:2000}%
  \BibitemOpen
  \bibfield  {author} {\bibinfo {author} {\bibfnamefont {T.}~\bibnamefont
  {Helgaker}}, \bibinfo {author} {\bibfnamefont {P.}~\bibnamefont {Jorgensen}},
  \ and\ \bibinfo {author} {\bibfnamefont {J.}~\bibnamefont {Olsen}},\
  }\href@noop {} {\emph {\bibinfo {title} {Molecular Electronic-structure
  Theory}}}\ (\bibinfo  {publisher} {Wiley},\ \bibinfo {year}
  {2000})\BibitemShut {NoStop}%
\bibitem [{\citenamefont {Bravyi}\ and\ \citenamefont
  {Kitaev}(2002)}]{Bravyi:2002/aph/210}%
  \BibitemOpen
  \bibfield  {author} {\bibinfo {author} {\bibfnamefont {S.~B.}\ \bibnamefont
  {Bravyi}}\ and\ \bibinfo {author} {\bibfnamefont {A.~Y.}\ \bibnamefont
  {Kitaev}},\ }\href {\doibase 10.1006/aphy.2002.6254} {\bibfield  {journal}
  {\bibinfo  {journal} {Ann. Phys.}\ }\textbf {\bibinfo {volume} {298}},\
  \bibinfo {pages} {210} (\bibinfo {year} {2002})}\BibitemShut {NoStop}%
\bibitem [{\citenamefont {Seeley}\ \emph {et~al.}(2012)\citenamefont {Seeley},
  \citenamefont {Richard},\ and\ \citenamefont
  {Love}}]{Seeley:2012/jcp/224109}%
  \BibitemOpen
  \bibfield  {author} {\bibinfo {author} {\bibfnamefont {J.~T.}\ \bibnamefont
  {Seeley}}, \bibinfo {author} {\bibfnamefont {M.~J.}\ \bibnamefont {Richard}},
  \ and\ \bibinfo {author} {\bibfnamefont {P.~J.}\ \bibnamefont {Love}},\
  }\href {\doibase 10.1063/1.4768229} {\bibfield  {journal} {\bibinfo
  {journal} {J. Chem. Phys.}\ }\textbf {\bibinfo {volume} {137}},\ \bibinfo
  {pages} {224109} (\bibinfo {year} {2012})}\BibitemShut {NoStop}%
\bibitem [{\citenamefont {Tranter}\ \emph {et~al.}(2015)\citenamefont
  {Tranter}, \citenamefont {Sofia}, \citenamefont {Seeley}, \citenamefont
  {Kaicher}, \citenamefont {McClean}, \citenamefont {Babbush}, \citenamefont
  {Coveney}, \citenamefont {Mintert}, \citenamefont {Wilhelm},\ and\
  \citenamefont {Love}}]{Tranter:2015/ijqc/1431}%
  \BibitemOpen
  \bibfield  {author} {\bibinfo {author} {\bibfnamefont {A.}~\bibnamefont
  {Tranter}}, \bibinfo {author} {\bibfnamefont {S.}~\bibnamefont {Sofia}},
  \bibinfo {author} {\bibfnamefont {J.}~\bibnamefont {Seeley}}, \bibinfo
  {author} {\bibfnamefont {M.}~\bibnamefont {Kaicher}}, \bibinfo {author}
  {\bibfnamefont {J.}~\bibnamefont {McClean}}, \bibinfo {author} {\bibfnamefont
  {R.}~\bibnamefont {Babbush}}, \bibinfo {author} {\bibfnamefont {P.~V.}\
  \bibnamefont {Coveney}}, \bibinfo {author} {\bibfnamefont {F.}~\bibnamefont
  {Mintert}}, \bibinfo {author} {\bibfnamefont {F.}~\bibnamefont {Wilhelm}}, \
  and\ \bibinfo {author} {\bibfnamefont {P.~J.}\ \bibnamefont {Love}},\ }\href
  {\doibase 10.1002/qua.24969} {\bibfield  {journal} {\bibinfo  {journal} {Int.
  J. Quantum Chem.}\ }\textbf {\bibinfo {volume} {115}},\ \bibinfo {pages}
  {1431} (\bibinfo {year} {2015})}\BibitemShut {NoStop}%
\bibitem [{\citenamefont {Setia}\ and\ \citenamefont
  {Whitfield}(2018)}]{Setia:2017/ArXiv/1712.00446}%
  \BibitemOpen
  \bibfield  {author} {\bibinfo {author} {\bibfnamefont {K.}~\bibnamefont
  {Setia}}\ and\ \bibinfo {author} {\bibfnamefont {J.~D.}\ \bibnamefont
  {Whitfield}},\ }\href {\doibase 10.1063/1.5019371} {\bibfield  {journal}
  {\bibinfo  {journal} {J. Chem. Phys.}\ }\textbf {\bibinfo {volume} {148}},\
  \bibinfo {pages} {164104} (\bibinfo {year} {2018})}\BibitemShut {NoStop}%
\bibitem [{\citenamefont {Havl{\'{i}}{\v{c}}ek}\ \emph
  {et~al.}(2017)\citenamefont {Havl{\'{i}}{\v{c}}ek}, \citenamefont {Troyer},\
  and\ \citenamefont {Whitfield}}]{Havlicek:2017/pra/032332}%
  \BibitemOpen
  \bibfield  {author} {\bibinfo {author} {\bibfnamefont {V.}~\bibnamefont
  {Havl{\'{i}}{\v{c}}ek}}, \bibinfo {author} {\bibfnamefont {M.}~\bibnamefont
  {Troyer}}, \ and\ \bibinfo {author} {\bibfnamefont {J.~D.}\ \bibnamefont
  {Whitfield}},\ }\href {\doibase 10.1103/PhysRevA.95.032332} {\bibfield
  {journal} {\bibinfo  {journal} {Phys. Rev. A}\ }\textbf {\bibinfo {volume}
  {95}},\ \bibinfo {pages} {032332} (\bibinfo {year} {2017})}\BibitemShut
  {NoStop}%
\bibitem [{\citenamefont {Cirac}\ and\ \citenamefont
  {Zoller}(2012)}]{cirac:2012}%
  \BibitemOpen
  \bibfield  {author} {\bibinfo {author} {\bibfnamefont {J.~I.}\ \bibnamefont
  {Cirac}}\ and\ \bibinfo {author} {\bibfnamefont {P.}~\bibnamefont {Zoller}},\
  }\href {\doibase 10.1038/nphys2275} {\bibfield  {journal} {\bibinfo
  {journal} {Nat. Phys.}\ }\textbf {\bibinfo {volume} {8}},\ \bibinfo {pages}
  {264} (\bibinfo {year} {2012})}\BibitemShut {NoStop}%
\bibitem [{\citenamefont {Argüello-Luengo}\ \emph {et~al.}(2018)\citenamefont
  {Argüello-Luengo}, \citenamefont {González-Tudela}, \citenamefont {Shi},
  \citenamefont {Zoller},\ and\ \citenamefont {Cirac}}]{cirac:2018}%
  \BibitemOpen
  \bibfield  {author} {\bibinfo {author} {\bibfnamefont {J.}~\bibnamefont
  {Argüello-Luengo}}, \bibinfo {author} {\bibfnamefont {A.}~\bibnamefont
  {González-Tudela}}, \bibinfo {author} {\bibfnamefont {T.}~\bibnamefont
  {Shi}}, \bibinfo {author} {\bibfnamefont {P.}~\bibnamefont {Zoller}}, \ and\
  \bibinfo {author} {\bibfnamefont {J.~I.}\ \bibnamefont {Cirac}},\ }\href@noop
  {} {\bibfield  {journal} {\bibinfo  {journal} {arXiv.org}\ ,\ \bibinfo
  {pages} {arXiv:1807.09228}} (\bibinfo {year} {2018})},\ \Eprint
  {http://arxiv.org/abs/1807.09228} {1807.09228} \BibitemShut {NoStop}%
\bibitem [{\citenamefont {Vogel}\ and\ \citenamefont
  {Risken}(1989)}]{PhysRevA.40.2847}%
  \BibitemOpen
  \bibfield  {author} {\bibinfo {author} {\bibfnamefont {K.}~\bibnamefont
  {Vogel}}\ and\ \bibinfo {author} {\bibfnamefont {H.}~\bibnamefont {Risken}},\
  }\href {\doibase 10.1103/PhysRevA.40.2847} {\bibfield  {journal} {\bibinfo
  {journal} {Phys. Rev. A}\ }\textbf {\bibinfo {volume} {40}},\ \bibinfo
  {pages} {2847} (\bibinfo {year} {1989})}\BibitemShut {NoStop}%
\bibitem [{\citenamefont {Cramer}\ \emph {et~al.}(2010)\citenamefont {Cramer},
  \citenamefont {Plenio}, \citenamefont {Flammia}, \citenamefont {Somma},
  \citenamefont {Gross}, \citenamefont {Bartlett}, \citenamefont
  {Landon-Cardinal}, \citenamefont {Poulin},\ and\ \citenamefont
  {Liu}}]{Cramer:2010bs}%
  \BibitemOpen
  \bibfield  {author} {\bibinfo {author} {\bibfnamefont {M.}~\bibnamefont
  {Cramer}}, \bibinfo {author} {\bibfnamefont {M.~B.}\ \bibnamefont {Plenio}},
  \bibinfo {author} {\bibfnamefont {S.~T.}\ \bibnamefont {Flammia}}, \bibinfo
  {author} {\bibfnamefont {R.}~\bibnamefont {Somma}}, \bibinfo {author}
  {\bibfnamefont {D.}~\bibnamefont {Gross}}, \bibinfo {author} {\bibfnamefont
  {S.~D.}\ \bibnamefont {Bartlett}}, \bibinfo {author} {\bibfnamefont
  {O.}~\bibnamefont {Landon-Cardinal}}, \bibinfo {author} {\bibfnamefont
  {D.}~\bibnamefont {Poulin}}, \ and\ \bibinfo {author} {\bibfnamefont {Y.-K.}\
  \bibnamefont {Liu}},\ }\href@noop {} {\bibfield  {journal} {\bibinfo
  {journal} {Nat. Commun.}\ }\textbf {\bibinfo {volume} {1}},\ \bibinfo {pages}
  {149} (\bibinfo {year} {2010})}\BibitemShut {NoStop}%
\bibitem [{\citenamefont {Verteletskyi}\ \emph {et~al.}(2019)\citenamefont
  {Verteletskyi}, \citenamefont {Yen},\ and\ \citenamefont
  {Izmaylov}}]{VVpaper1}%
  \BibitemOpen
  \bibfield  {author} {\bibinfo {author} {\bibfnamefont {V.}~\bibnamefont
  {Verteletskyi}}, \bibinfo {author} {\bibfnamefont {T.-C.}\ \bibnamefont
  {Yen}}, \ and\ \bibinfo {author} {\bibfnamefont {A.~F.}\ \bibnamefont
  {Izmaylov}},\ }\href@noop {} {\bibfield  {journal} {\bibinfo  {journal}
  {arXiv.org}\ ,\ \bibinfo {pages} {arXiv:1907.03358}} (\bibinfo {year}
  {2019})},\ \Eprint {http://arxiv.org/abs/1907.03358} {1907.03358}
  \BibitemShut {NoStop}%
\bibitem [{\citenamefont {Izmaylov}\ \emph {et~al.}(2019)\citenamefont
  {Izmaylov}, \citenamefont {Yen},\ and\ \citenamefont
  {Ryabinkin}}]{Izmaylov:2019gb}%
  \BibitemOpen
  \bibfield  {author} {\bibinfo {author} {\bibfnamefont {A.~F.}\ \bibnamefont
  {Izmaylov}}, \bibinfo {author} {\bibfnamefont {T.-C.}\ \bibnamefont {Yen}}, \
  and\ \bibinfo {author} {\bibfnamefont {I.~G.}\ \bibnamefont {Ryabinkin}},\
  }\href@noop {} {\bibfield  {journal} {\bibinfo  {journal} {Chem. Sci.}\
  }\textbf {\bibinfo {volume} {10}},\ \bibinfo {pages} {3746} (\bibinfo {year}
  {2019})}\BibitemShut {NoStop}%
\bibitem [{\citenamefont {Albarr{\'a}n-Arriagada}\ \emph
  {et~al.}(2018)\citenamefont {Albarr{\'a}n-Arriagada}, \citenamefont
  {Barrios}, \citenamefont {Sanz}, \citenamefont {Romero}, \citenamefont
  {Lamata}, \citenamefont {Retamal},\ and\ \citenamefont
  {Solano}}]{nArriagada:2018ju}%
  \BibitemOpen
  \bibfield  {author} {\bibinfo {author} {\bibfnamefont {F.}~\bibnamefont
  {Albarr{\'a}n-Arriagada}}, \bibinfo {author} {\bibfnamefont {G.~A.}\
  \bibnamefont {Barrios}}, \bibinfo {author} {\bibfnamefont {M.}~\bibnamefont
  {Sanz}}, \bibinfo {author} {\bibfnamefont {G.}~\bibnamefont {Romero}},
  \bibinfo {author} {\bibfnamefont {L.}~\bibnamefont {Lamata}}, \bibinfo
  {author} {\bibfnamefont {J.~C.}\ \bibnamefont {Retamal}}, \ and\ \bibinfo
  {author} {\bibfnamefont {E.}~\bibnamefont {Solano}},\ }\href@noop {}
  {\bibfield  {journal} {\bibinfo  {journal} {Phys. Rev. A}\ }\textbf {\bibinfo
  {volume} {97}},\ \bibinfo {pages} {032320:1} (\bibinfo {year}
  {2018})}\BibitemShut {NoStop}%
\bibitem [{\citenamefont {Prevedel}\ \emph {et~al.}(2007)\citenamefont
  {Prevedel}, \citenamefont {Walther}, \citenamefont {Tiefenbacher},
  \citenamefont {B{\"o}hi}, \citenamefont {Kaltenbaek}, \citenamefont
  {Jennewein},\ and\ \citenamefont {Zeilinger}}]{Prevedel:2007ca}%
  \BibitemOpen
  \bibfield  {author} {\bibinfo {author} {\bibfnamefont {R.}~\bibnamefont
  {Prevedel}}, \bibinfo {author} {\bibfnamefont {P.}~\bibnamefont {Walther}},
  \bibinfo {author} {\bibfnamefont {F.}~\bibnamefont {Tiefenbacher}}, \bibinfo
  {author} {\bibfnamefont {P.}~\bibnamefont {B{\"o}hi}}, \bibinfo {author}
  {\bibfnamefont {R.}~\bibnamefont {Kaltenbaek}}, \bibinfo {author}
  {\bibfnamefont {T.}~\bibnamefont {Jennewein}}, \ and\ \bibinfo {author}
  {\bibfnamefont {A.}~\bibnamefont {Zeilinger}},\ }\href@noop {} {\bibfield
  {journal} {\bibinfo  {journal} {Nature}\ }\textbf {\bibinfo {volume} {445}},\
  \bibinfo {pages} {65} (\bibinfo {year} {2007})}\BibitemShut {NoStop}%
\bibitem [{\citenamefont {Procopio}\ \emph {et~al.}(2015)\citenamefont
  {Procopio}, \citenamefont {Moqanaki}, \citenamefont {Ara{\'u}jo},
  \citenamefont {Costa}, \citenamefont {Calafell}, \citenamefont {Dowd},
  \citenamefont {Hamel}, \citenamefont {Rozema}, \citenamefont {Brukner},\ and\
  \citenamefont {Walther}}]{Moqanaki:2015iw}%
  \BibitemOpen
  \bibfield  {author} {\bibinfo {author} {\bibfnamefont {L.~M.}\ \bibnamefont
  {Procopio}}, \bibinfo {author} {\bibfnamefont {A.}~\bibnamefont {Moqanaki}},
  \bibinfo {author} {\bibfnamefont {M.}~\bibnamefont {Ara{\'u}jo}}, \bibinfo
  {author} {\bibfnamefont {F.}~\bibnamefont {Costa}}, \bibinfo {author}
  {\bibfnamefont {I.~A.}\ \bibnamefont {Calafell}}, \bibinfo {author}
  {\bibfnamefont {E.~G.}\ \bibnamefont {Dowd}}, \bibinfo {author}
  {\bibfnamefont {D.~R.}\ \bibnamefont {Hamel}}, \bibinfo {author}
  {\bibfnamefont {L.~A.}\ \bibnamefont {Rozema}}, \bibinfo {author}
  {\bibfnamefont {v.}~\bibnamefont {Brukner}}, \ and\ \bibinfo {author}
  {\bibfnamefont {P.}~\bibnamefont {Walther}},\ }\href@noop {} {\bibfield
  {journal} {\bibinfo  {journal} {Nature Communications}\ }\textbf {\bibinfo
  {volume} {6}},\ \bibinfo {pages} {7913:1} (\bibinfo {year}
  {2015})}\BibitemShut {NoStop}%
\bibitem [{\citenamefont {Reimer}\ \emph {et~al.}(2019)\citenamefont {Reimer},
  \citenamefont {Sciara}, \citenamefont {Roztocki}, \citenamefont {Islam},
  \citenamefont {Cort{\'e}s}, \citenamefont {Zhang}, \citenamefont {Fischer},
  \citenamefont {Loranger}, \citenamefont {Kashyap}, \citenamefont {Cino},
  \citenamefont {Chu}, \citenamefont {Little}, \citenamefont {Moss},
  \citenamefont {Caspani}, \citenamefont {Munro}, \citenamefont {Aza{\~n}a},
  \citenamefont {Kues},\ and\ \citenamefont {Morandotti}}]{Reimer:2018cv}%
  \BibitemOpen
  \bibfield  {author} {\bibinfo {author} {\bibfnamefont {C.}~\bibnamefont
  {Reimer}}, \bibinfo {author} {\bibfnamefont {S.}~\bibnamefont {Sciara}},
  \bibinfo {author} {\bibfnamefont {P.}~\bibnamefont {Roztocki}}, \bibinfo
  {author} {\bibfnamefont {M.}~\bibnamefont {Islam}}, \bibinfo {author}
  {\bibfnamefont {L.~R.}\ \bibnamefont {Cort{\'e}s}}, \bibinfo {author}
  {\bibfnamefont {Y.}~\bibnamefont {Zhang}}, \bibinfo {author} {\bibfnamefont
  {B.}~\bibnamefont {Fischer}}, \bibinfo {author} {\bibfnamefont
  {S.}~\bibnamefont {Loranger}}, \bibinfo {author} {\bibfnamefont
  {R.}~\bibnamefont {Kashyap}}, \bibinfo {author} {\bibfnamefont
  {A.}~\bibnamefont {Cino}}, \bibinfo {author} {\bibfnamefont {S.~T.}\
  \bibnamefont {Chu}}, \bibinfo {author} {\bibfnamefont {B.~E.}\ \bibnamefont
  {Little}}, \bibinfo {author} {\bibfnamefont {D.~J.}\ \bibnamefont {Moss}},
  \bibinfo {author} {\bibfnamefont {L.}~\bibnamefont {Caspani}}, \bibinfo
  {author} {\bibfnamefont {W.~J.}\ \bibnamefont {Munro}}, \bibinfo {author}
  {\bibfnamefont {J.}~\bibnamefont {Aza{\~n}a}}, \bibinfo {author}
  {\bibfnamefont {M.}~\bibnamefont {Kues}}, \ and\ \bibinfo {author}
  {\bibfnamefont {R.}~\bibnamefont {Morandotti}},\ }\href@noop {} {\bibfield
  {journal} {\bibinfo  {journal} {Nature Physics}\ }\textbf {\bibinfo {volume}
  {15}},\ \bibinfo {pages} {148} (\bibinfo {year} {2019})}\BibitemShut
  {NoStop}%
\bibitem [{\citenamefont {Bravyi}\ \emph {et~al.}(2017)\citenamefont {Bravyi},
  \citenamefont {Gambetta}, \citenamefont {Mezzacapo},\ and\ \citenamefont
  {Temme}}]{Bravyi:2017wb}%
  \BibitemOpen
  \bibfield  {author} {\bibinfo {author} {\bibfnamefont {S.}~\bibnamefont
  {Bravyi}}, \bibinfo {author} {\bibfnamefont {J.~M.}\ \bibnamefont
  {Gambetta}}, \bibinfo {author} {\bibfnamefont {A.}~\bibnamefont {Mezzacapo}},
  \ and\ \bibinfo {author} {\bibfnamefont {K.}~\bibnamefont {Temme}},\
  }\href@noop {} {\bibfield  {journal} {\bibinfo  {journal} {arXiv.org}\ ,\
  \bibinfo {pages} {arXiv:1701.08213}} (\bibinfo {year} {2017})},\ \Eprint
  {http://arxiv.org/abs/1701.08213v1} {1701.08213v1} \BibitemShut {NoStop}%
\bibitem [{\citenamefont {Nielsen}\ and\ \citenamefont
  {Chuang}(2010)}]{Nielsen:2010}%
  \BibitemOpen
  \bibfield  {author} {\bibinfo {author} {\bibfnamefont {M.}~\bibnamefont
  {Nielsen}}\ and\ \bibinfo {author} {\bibfnamefont {I.}~\bibnamefont
  {Chuang}},\ }\href {https://books.google.ca/books?id=j2ULnwEACAAJ} {\emph
  {\bibinfo {title} {{Quantum Computation and Quantum Information: 10th
  Anniversary Edition}}}}\ (\bibinfo  {publisher} {Cambridge University
  Press},\ \bibinfo {year} {2010})\BibitemShut {NoStop}%
\bibitem [{\citenamefont {{Rigetti\ Computing}}(2018)}]{Rigetti_doc}%
  \BibitemOpen
  \bibfield  {author} {\bibinfo {author} {\bibnamefont {{Rigetti\
  Computing}}},\ }\href@noop {} {\enquote {\bibinfo {title} {{pyQuil 1.9}},}\ }
  (\bibinfo {year} {2018}),\ \bibinfo {note}
  {http://docs.rigetti.com/en/1.9/qpu.html}\BibitemShut {NoStop}%
\bibitem [{\citenamefont {Welsh}(1967)}]{Welsh:1967}%
  \BibitemOpen
  \bibfield  {author} {\bibinfo {author} {\bibfnamefont {D.~J.~A.}\
  \bibnamefont {Welsh}},\ }\href {\doibase 10.1093/comjnl/10.1.85} {\bibfield
  {journal} {\bibinfo  {journal} {Comput. J.}\ }\textbf {\bibinfo {volume}
  {10}},\ \bibinfo {pages} {85–86} (\bibinfo {year} {1967})}\BibitemShut
  {NoStop}%
\bibitem [{\citenamefont {Matula}\ \emph {et~al.}(1972)\citenamefont {Matula},
  \citenamefont {Marble},\ and\ \citenamefont {Isaacson}}]{Matula:1972}%
  \BibitemOpen
  \bibfield  {author} {\bibinfo {author} {\bibfnamefont {D.~W.}\ \bibnamefont
  {Matula}}, \bibinfo {author} {\bibfnamefont {G.}~\bibnamefont {Marble}}, \
  and\ \bibinfo {author} {\bibfnamefont {J.~D.}\ \bibnamefont {Isaacson}},\
  }in\ \href {\doibase 10.1016/b978-1-4832-3187-7.50015-5} {\emph {\bibinfo
  {booktitle} {Graph Theory and Computing}}},\ \bibinfo {editor} {edited by\
  \bibinfo {editor} {\bibfnamefont {R.~C.}\ \bibnamefont {Read}}}\ (\bibinfo
  {publisher} {Academic Press},\ \bibinfo {year} {1972})\ pp.\ \bibinfo {pages}
  {109 -- 122}\BibitemShut {NoStop}%
\bibitem [{\citenamefont {Brélaz}(1979)}]{Brelaz:1973}%
  \BibitemOpen
  \bibfield  {author} {\bibinfo {author} {\bibfnamefont {D.}~\bibnamefont
  {Brélaz}},\ }\href {\doibase 10.1145/359094.359101} {\bibfield  {journal}
  {\bibinfo  {journal} {Commun. ACM}\ }\textbf {\bibinfo {volume} {22}},\
  \bibinfo {pages} {251–256} (\bibinfo {year} {1979})}\BibitemShut {NoStop}%
\bibitem [{\citenamefont {Leighton}(1979)}]{Leighton:1979}%
  \BibitemOpen
  \bibfield  {author} {\bibinfo {author} {\bibfnamefont {F.~T.}\ \bibnamefont
  {Leighton}},\ }\href@noop {} {\bibfield  {journal} {\bibinfo  {journal} {J.
  Res. Natl. Bur. Stand.}\ }\textbf {\bibinfo {volume} {84}},\ \bibinfo {pages}
  {489} (\bibinfo {year} {1979})}\BibitemShut {NoStop}%
\bibitem [{\citenamefont {Dutton}\ and\ \citenamefont
  {Brigham}(1981)}]{dutton_brigham_1981}%
  \BibitemOpen
  \bibfield  {author} {\bibinfo {author} {\bibfnamefont {R.~D.}\ \bibnamefont
  {Dutton}}\ and\ \bibinfo {author} {\bibfnamefont {R.~C.}\ \bibnamefont
  {Brigham}},\ }\href {\doibase 10.1093/comjnl/24.1.85} {\bibfield  {journal}
  {\bibinfo  {journal} {Comput. J.}\ }\textbf {\bibinfo {volume} {24}},\
  \bibinfo {pages} {85–86} (\bibinfo {year} {1981})}\BibitemShut {NoStop}%
\bibitem [{\citenamefont {Hertz}(1990)}]{hertz_1990}%
  \BibitemOpen
  \bibfield  {author} {\bibinfo {author} {\bibfnamefont {A.}~\bibnamefont
  {Hertz}},\ }\href {\doibase 10.1016/0095-8956(90)90078-e} {\bibfield
  {journal} {\bibinfo  {journal} {J. Comb. Theory}\ }\textbf {\bibinfo {volume}
  {50}},\ \bibinfo {pages} {231–240} (\bibinfo {year} {1990})}\BibitemShut
  {NoStop}%
\bibitem [{\citenamefont {Boppana}\ and\ \citenamefont
  {Halldórsson}(1992)}]{boppana}%
  \BibitemOpen
  \bibfield  {author} {\bibinfo {author} {\bibfnamefont {R.}~\bibnamefont
  {Boppana}}\ and\ \bibinfo {author} {\bibfnamefont {M.~M.}\ \bibnamefont
  {Halldórsson}},\ }\href {\doibase 10.1007/bf01994876} {\bibfield  {journal}
  {\bibinfo  {journal} {BIT Numer. Math}\ }\textbf {\bibinfo {volume} {32}},\
  \bibinfo {pages} {180–196} (\bibinfo {year} {1992})}\BibitemShut {NoStop}%
\bibitem [{\citenamefont {Tomita}\ \emph {et~al.}(2006)\citenamefont {Tomita},
  \citenamefont {Tanaka},\ and\ \citenamefont
  {Takahashi}}]{tomita_tanaka_takahashi_2006}%
  \BibitemOpen
  \bibfield  {author} {\bibinfo {author} {\bibfnamefont {E.}~\bibnamefont
  {Tomita}}, \bibinfo {author} {\bibfnamefont {A.}~\bibnamefont {Tanaka}}, \
  and\ \bibinfo {author} {\bibfnamefont {H.}~\bibnamefont {Takahashi}},\ }\href
  {\doibase 10.1016/j.tcs.2006.06.015} {\bibfield  {journal} {\bibinfo
  {journal} {Theor. Comput. Sci.}\ }\textbf {\bibinfo {volume} {363}},\
  \bibinfo {pages} {28–42} (\bibinfo {year} {2006})}\BibitemShut {NoStop}%
\bibitem [{\citenamefont {Aaronson}\ and\ \citenamefont
  {Gottesman}(2004)}]{Gottesman:CG}%
  \BibitemOpen
  \bibfield  {author} {\bibinfo {author} {\bibfnamefont {S.}~\bibnamefont
  {Aaronson}}\ and\ \bibinfo {author} {\bibfnamefont {D.}~\bibnamefont
  {Gottesman}},\ }\href {\doibase 10.1103/PhysRevA.70.052328} {\bibfield
  {journal} {\bibinfo  {journal} {Phys. Rev. A}\ }\textbf {\bibinfo {volume}
  {70}},\ \bibinfo {pages} {052328} (\bibinfo {year} {2004})}\BibitemShut
  {NoStop}%
\bibitem [{\citenamefont {Wootters}\ and\ \citenamefont
  {Fields}(1989)}]{MUB_QST1}%
  \BibitemOpen
  \bibfield  {author} {\bibinfo {author} {\bibfnamefont {W.}~\bibnamefont
  {Wootters}}\ and\ \bibinfo {author} {\bibfnamefont {B.}~\bibnamefont
  {Fields}},\ }\href@noop {} {\bibfield  {journal} {\bibinfo  {journal} {Ann.
  Phys. (N.Y.)}\ }\textbf {\bibinfo {volume} {191}},\ \bibinfo {pages} {363}
  (\bibinfo {year} {1989})}\BibitemShut {NoStop}%
\bibitem [{\citenamefont {Sainz}\ \emph {et~al.}(2018)\citenamefont {Sainz},
  \citenamefont {Garcia},\ and\ \citenamefont {Klimov}}]{MUB_QST2}%
  \BibitemOpen
  \bibfield  {author} {\bibinfo {author} {\bibfnamefont {I.}~\bibnamefont
  {Sainz}}, \bibinfo {author} {\bibfnamefont {A.}~\bibnamefont {Garcia}}, \
  and\ \bibinfo {author} {\bibfnamefont {A.~B.}\ \bibnamefont {Klimov}},\
  }\href@noop {} {\bibfield  {journal} {\bibinfo  {journal} {Phys. Lett. A}\
  }\textbf {\bibinfo {volume} {382}},\ \bibinfo {pages} {66} (\bibinfo {year}
  {2018})}\BibitemShut {NoStop}%
\bibitem [{\citenamefont {Altepeter}\ \emph {et~al.}(2004)\citenamefont
  {Altepeter}, \citenamefont {James},\ and\ \citenamefont {Kwiat}}]{DJ_2004}%
  \BibitemOpen
  \bibfield  {author} {\bibinfo {author} {\bibfnamefont {J.~B.}\ \bibnamefont
  {Altepeter}}, \bibinfo {author} {\bibfnamefont {D.~F.}\ \bibnamefont
  {James}}, \ and\ \bibinfo {author} {\bibfnamefont {P.~G.}\ \bibnamefont
  {Kwiat}},\ }in\ \href@noop {} {\emph {\bibinfo {booktitle} {Lect. Notes
  Phys.: Quantum State Estimation}}},\ Vol.\ \bibinfo {volume} {649}\ (\bibinfo
  {year} {2004})\ pp.\ \bibinfo {pages} {113--145}\BibitemShut {NoStop}%
\bibitem [{\citenamefont {Lawrence}\ \emph {et~al.}(2002)\citenamefont
  {Lawrence}, \citenamefont {Brukner},\ and\ \citenamefont
  {Zeilinger}}]{Zeilenger:PRA/2002}%
  \BibitemOpen
  \bibfield  {author} {\bibinfo {author} {\bibfnamefont {J.}~\bibnamefont
  {Lawrence}}, \bibinfo {author} {\bibfnamefont {C.}~\bibnamefont {Brukner}}, \
  and\ \bibinfo {author} {\bibfnamefont {A.}~\bibnamefont {Zeilinger}},\ }\href
  {\doibase 10.1103/PhysRevA.65.032320} {\bibfield  {journal} {\bibinfo
  {journal} {Phys. Rev. A}\ }\textbf {\bibinfo {volume} {65}},\ \bibinfo
  {pages} {032320} (\bibinfo {year} {2002})}\BibitemShut {NoStop}%
\bibitem [{\citenamefont {Jena}\ \emph {et~al.}(2019)\citenamefont {Jena},
  \citenamefont {Genin},\ and\ \citenamefont {Mosca}}]{MoscaA}%
  \BibitemOpen
  \bibfield  {author} {\bibinfo {author} {\bibfnamefont {A.}~\bibnamefont
  {Jena}}, \bibinfo {author} {\bibfnamefont {S.}~\bibnamefont {Genin}}, \ and\
  \bibinfo {author} {\bibfnamefont {M.}~\bibnamefont {Mosca}},\ }\href@noop {}
  {\bibfield  {journal} {\bibinfo  {journal} {arXiv.org}\ ,\ \bibinfo {pages}
  {arXiv:1907.07859}} (\bibinfo {year} {2019})},\ \Eprint
  {http://arxiv.org/abs/1907.07859} {1907.07859} \BibitemShut {NoStop}%
\bibitem [{\citenamefont {Izmaylov}\ \emph {et~al.}(2020)\citenamefont
  {Izmaylov}, \citenamefont {Yen}, \citenamefont {Lang},\ and\ \citenamefont
  {Verteletskyi}}]{IzmaylovA}%
  \BibitemOpen
  \bibfield  {author} {\bibinfo {author} {\bibfnamefont {A.~F.}\ \bibnamefont
  {Izmaylov}}, \bibinfo {author} {\bibfnamefont {T.-C.}\ \bibnamefont {Yen}},
  \bibinfo {author} {\bibfnamefont {R.~A.}\ \bibnamefont {Lang}}, \ and\
  \bibinfo {author} {\bibfnamefont {V.}~\bibnamefont {Verteletskyi}},\
  }\href@noop {} {\bibfield  {journal} {\bibinfo  {journal} {J. Chem. Theory
  Comput.}\ }\textbf {\bibinfo {volume} {16}},\ \bibinfo {pages} {190}
  (\bibinfo {year} {2020})}\BibitemShut {NoStop}%
\bibitem [{\citenamefont {Huggins}\ \emph {et~al.}(2019)\citenamefont
  {Huggins}, \citenamefont {McClean}, \citenamefont {Rubin}, \citenamefont
  {Jiang}, \citenamefont {Wiebe}, \citenamefont {Whaley},\ and\ \citenamefont
  {Babbush}}]{BabbushA}%
  \BibitemOpen
  \bibfield  {author} {\bibinfo {author} {\bibfnamefont {W.~J.}\ \bibnamefont
  {Huggins}}, \bibinfo {author} {\bibfnamefont {J.}~\bibnamefont {McClean}},
  \bibinfo {author} {\bibfnamefont {N.}~\bibnamefont {Rubin}}, \bibinfo
  {author} {\bibfnamefont {Z.}~\bibnamefont {Jiang}}, \bibinfo {author}
  {\bibfnamefont {N.}~\bibnamefont {Wiebe}}, \bibinfo {author} {\bibfnamefont
  {K.~B.}\ \bibnamefont {Whaley}}, \ and\ \bibinfo {author} {\bibfnamefont
  {R.}~\bibnamefont {Babbush}},\ }\href@noop {} {\bibfield  {journal} {\bibinfo
   {journal} {arXiv.org}\ ,\ \bibinfo {pages} {arXiv:1907.13117}} (\bibinfo
  {year} {2019})},\ \Eprint {http://arxiv.org/abs/1907.13117} {1907.13117}
  \BibitemShut {NoStop}%
\bibitem [{\citenamefont {Gokhale}\ \emph {et~al.}(2019)\citenamefont
  {Gokhale}, \citenamefont {Angiuli}, \citenamefont {Ding}, \citenamefont
  {Gui}, \citenamefont {Tomesh}, \citenamefont {Suchara}, \citenamefont
  {Martonosi},\ and\ \citenamefont {Chong}}]{ChicagoA}%
  \BibitemOpen
  \bibfield  {author} {\bibinfo {author} {\bibfnamefont {P.}~\bibnamefont
  {Gokhale}}, \bibinfo {author} {\bibfnamefont {O.}~\bibnamefont {Angiuli}},
  \bibinfo {author} {\bibfnamefont {Y.}~\bibnamefont {Ding}}, \bibinfo {author}
  {\bibfnamefont {K.}~\bibnamefont {Gui}}, \bibinfo {author} {\bibfnamefont
  {T.}~\bibnamefont {Tomesh}}, \bibinfo {author} {\bibfnamefont
  {M.}~\bibnamefont {Suchara}}, \bibinfo {author} {\bibfnamefont
  {M.}~\bibnamefont {Martonosi}}, \ and\ \bibinfo {author} {\bibfnamefont
  {F.~T.}\ \bibnamefont {Chong}},\ }\href@noop {} {\bibfield  {journal}
  {\bibinfo  {journal} {arXiv.org}\ ,\ \bibinfo {pages} {arXiv:1907.13623}}
  (\bibinfo {year} {2019})},\ \Eprint {http://arxiv.org/abs/1907.13623}
  {1907.13623} \BibitemShut {NoStop}%
\bibitem [{\citenamefont {Zhao}\ \emph {et~al.}(2019)\citenamefont {Zhao},
  \citenamefont {Tranter}, \citenamefont {Kirby}, \citenamefont {Ung},
  \citenamefont {Miyake},\ and\ \citenamefont {Love}}]{Zhao:2019vz}%
  \BibitemOpen
  \bibfield  {author} {\bibinfo {author} {\bibfnamefont {A.}~\bibnamefont
  {Zhao}}, \bibinfo {author} {\bibfnamefont {A.}~\bibnamefont {Tranter}},
  \bibinfo {author} {\bibfnamefont {W.~M.}\ \bibnamefont {Kirby}}, \bibinfo
  {author} {\bibfnamefont {S.~F.}\ \bibnamefont {Ung}}, \bibinfo {author}
  {\bibfnamefont {A.}~\bibnamefont {Miyake}}, \ and\ \bibinfo {author}
  {\bibfnamefont {P.}~\bibnamefont {Love}},\ }\href@noop {} {\bibfield
  {journal} {\bibinfo  {journal} {arXiv.org}\ ,\ \bibinfo {pages}
  {arXiv:1908.08067v1}} (\bibinfo {year} {2019})}\BibitemShut {NoStop}%
\bibitem [{\citenamefont {Bonet-Monroig}\ \emph {et~al.}(2019)\citenamefont
  {Bonet-Monroig}, \citenamefont {Babbush},\ and\ \citenamefont
  {O'Brien}}]{BonetMonroig:2019wv}%
  \BibitemOpen
  \bibfield  {author} {\bibinfo {author} {\bibfnamefont {X.}~\bibnamefont
  {Bonet-Monroig}}, \bibinfo {author} {\bibfnamefont {R.}~\bibnamefont
  {Babbush}}, \ and\ \bibinfo {author} {\bibfnamefont {T.~E.}\ \bibnamefont
  {O'Brien}},\ }\href@noop {} {\bibfield  {journal} {\bibinfo  {journal}
  {arXiv.org}\ } (\bibinfo {year} {2019})},\ \Eprint
  {http://arxiv.org/abs/1908.05628v1} {1908.05628v1} \BibitemShut {NoStop}%
\bibitem [{\citenamefont {Crawford}\ \emph {et~al.}(2019)\citenamefont
  {Crawford}, \citenamefont {van Straaten}, \citenamefont {Wang}, \citenamefont
  {Parks}, \citenamefont {Campbell},\ and\ \citenamefont
  {Brierley}}]{Crawford:2019tg}%
  \BibitemOpen
  \bibfield  {author} {\bibinfo {author} {\bibfnamefont {O.}~\bibnamefont
  {Crawford}}, \bibinfo {author} {\bibfnamefont {B.}~\bibnamefont {van
  Straaten}}, \bibinfo {author} {\bibfnamefont {D.}~\bibnamefont {Wang}},
  \bibinfo {author} {\bibfnamefont {T.}~\bibnamefont {Parks}}, \bibinfo
  {author} {\bibfnamefont {E.}~\bibnamefont {Campbell}}, \ and\ \bibinfo
  {author} {\bibfnamefont {S.}~\bibnamefont {Brierley}},\ }\href@noop {}
  {\bibfield  {journal} {\bibinfo  {journal} {arXiv.org}\ } (\bibinfo {year}
  {2019})},\ \Eprint {http://arxiv.org/abs/1908.06942v1} {1908.06942v1}
  \BibitemShut {NoStop}%
\bibitem [{\citenamefont {Marle}\ and\ \citenamefont
  {Libermann}(1987)}]{SympGeomBook}%
  \BibitemOpen
  \bibfield  {author} {\bibinfo {author} {\bibfnamefont {C.}~\bibnamefont
  {Marle}}\ and\ \bibinfo {author} {\bibfnamefont {P.}~\bibnamefont
  {Libermann}},\ }in\ \href@noop {} {\emph {\bibinfo {booktitle} {Symplectic
  geometry and analytical mechanics}}}\ (\bibinfo {year} {1987})\ p.~\bibinfo
  {pages} {15}\BibitemShut {NoStop}%
\bibitem [{\citenamefont {Ryabinkin}\ \emph {et~al.}(2018)\citenamefont
  {Ryabinkin}, \citenamefont {Yen}, \citenamefont {Genin},\ and\ \citenamefont
  {Izmaylov}}]{Ryabinkin:2018/qcc}%
  \BibitemOpen
  \bibfield  {author} {\bibinfo {author} {\bibfnamefont {I.~G.}\ \bibnamefont
  {Ryabinkin}}, \bibinfo {author} {\bibfnamefont {T.-C.}\ \bibnamefont {Yen}},
  \bibinfo {author} {\bibfnamefont {S.~N.}\ \bibnamefont {Genin}}, \ and\
  \bibinfo {author} {\bibfnamefont {A.~F.}\ \bibnamefont {Izmaylov}},\
  }\href@noop {} {\bibfield  {journal} {\bibinfo  {journal} {J. Chem. Theory
  Comput.}\ }\textbf {\bibinfo {volume} {14}},\ \bibinfo {pages} {6317}
  (\bibinfo {year} {2018})}\BibitemShut {NoStop}%
\end{thebibliography}

\end{document}